%% file: 47Tuc_paper2_v8.tex
\documentclass{aa}
\usepackage{graphicx}
\usepackage{txfonts}
\usepackage{epsf}  
\usepackage{amssymb}  
\usepackage{rotating}
\usepackage{url}
\usepackage{lscape} 
\usepackage{subfig}
\usepackage{subfloat}
\usepackage{longtable}
\usepackage{natbib}
\usepackage{amsmath} 
\newcommand{\angstrom}{\textup{\AA}}
\bibpunct{(}{)}{;}{a}{}{,} 
\begin{document}
%
%
\newcommand{\ie}{i.e.}
\newcommand{\eg}{e.g.}
\newcommand{\cf}{cf.}	
\newcommand{\kms}{km\,s$^{-1}$}
\newcommand{\teff}{$T_{\rm eff}$}
\newcommand{\logg}{$\log g$}
\newcommand{\feh}{[Fe/H]}
\newcommand{\msun}{${\rm M}_{\odot}$}
\newcommand{\percent}{\,{\%}}
\newcommand{\vmic}{$\xi_t$}
\newcommand{\vsini}{$v \sin i$}
\newcommand{\feone}{Fe\,{\sc I}}
\newcommand{\fetwo}{Fe\,{\sc II}}
\newcommand{\loggf}{log$(gf)$}
\newcommand{\mghi}{$^{24}$MgH}
\newcommand{\mghii}{$^{25}$MgH}
\newcommand{\mghiii}{$^{26}$MgH}
\newcommand{\mgi}{$^{24}$Mg}
\newcommand{\mgii}{$^{25}$Mg}
\newcommand{\mgiii}{$^{26}$Mg}

\title{The chemical composition of red giants in 47 Tucanae. II. Magnesium isotopes and pollution scenarios
\thanks{Based on observations made with the ESO Very Large Telescope at Paranal Observatory, Chile (Programmes 084.B-0810 and 086.B-0237).}}
\titlerunning{The chemical composition of 47Tuc II}
\authorrunning{A. O. Thygesen et al.}
\author{
A. O.~Thygesen\inst{1,2} 
\and
L.~Sbordone\inst{3,4,2}
\and
H.-G.~Ludwig\inst{2}
\and
P.~Ventura\inst{5}
\and
D.~Yong\inst{6}
\and
R.~Collet\inst{6}
\and
N.~Christlieb\inst{2}
\and
J.~Melendez\inst{7}
\and
S.~Zaggia\inst{8}
} 
\offprints{A. O.~Thygesen}
\mail{aot@astro.caltech.edu}
\institute{California Institute of Technology, 1200 E. California Blvd., MC 249-17, Pasadena, CA 91125, USA.
\and
Zentrum f\"{u}r Astronomie der Universit\"{a}t Heidelberg, Landessternwarte, K\"{o}nigstuhl 12, 69117 Heidelberg, Germany.
\and
Millennium Institute of Astrophysics, Chile.
\and
Pontificia Universidad Cat{\'o}lica de Chile, Av. Vicu{\~n}a Mackenna 4860, 782-0436 Macul, Santiago, Chile.
\and
INAF-Osservatorio Astronomico di Roma, Via Frascati 33, I-00040 Monte Porzio Catone, Italy.
\and
Research School of Astronomy and Astrophysics, Australian National University, Canberra, ACT 0200, Australia.
\and
Departamento de Astronomia do IAG/USP, Universidade de S\~{a}o Paulo, rua do M\~{a}tao 1226, 05508-900, S\~{a}o Paulo, SP, Brasil.
\and
INAF-Osservatorio Astronomico di Padova, Vicolo dell'Osservatorio 5, 35122 Padova, Italy.
}
\date{Received 01-06 2015 ; Accepted 25-01 2016 }
\abstract
{The phenomenon of multiple populations in globular clusters is still far from understood, with several proposed mechanisms to explain the observed behaviour. The study of elemental and isotopic abundance patterns are crucial for investigating the differences among candidate pollution mechanisms.} 
{We derive magnesium isotopic ratios for 13 stars in the globular cluster 47 Tucanae (NGC 104) to provide new, detailed information about the nucleosynthesis that has occurred within the cluster. For the first time, the impact of 3D model stellar atmospheres on the derived Mg isotopic ratios is investigated.}
{Using both tailored 1D atmospheric models and 3D hydrodynamical models, we derive magnesium isotopic ratios from four features of MgH near 5135\,\AA\ in 13 giants near the tip of the RGB, using high signal-to-noise, high-resolution spectra.}
{We derive the magnesium isotopic ratios for all stars and find no significant offset of the isotopic distribution between the pristine and the polluted populations. Furthermore, we do not detect any statistically significant differences in the spread in the Mg isotopes in either population. No trends were found between the Mg isotopes and [Al/Fe]. The inclusion of 3D atmospheres has a significant impact on the derived \mgii/\mgi\ ratio, increasing it by a factor of up to 2.5, compared to 1D. The \mgiii/\mgi\ ratio, on the other hand, essentially remains unchanged.}
{We confirm the results seen from other globular clusters, where no strong variation in the isotopic ratios is observed between stellar populations, for observed ranges in [Al/Fe]. We see no evidence for any significant activation of the Mg-Al burning chain. The use of 3D atmospheres causes an increase of a factor of up to 2.5 in the fraction of \mgii, resolving part of the discrepancy between the observed isotopic fraction and the predictions from pollution models.}
\maketitle

\section{Introduction}
The multiple population phenomenon in globular clusters (GCs) is at the present not understood well. Essentially, all well-studied GCs are found to harbour a \emph{pristine} population of stars with a composition similar to field stars at the same metallicity, and one or more \emph{polluted} populations, which show anomalies in the light element abundances. The only known exception is the GC Ruprecht 106 \citep{villanova}, which appears to be a single stellar population. Whereas it is observationally well-established that most, if not all, GCs show variations in the light elements (He,	C, N, O, Na, sometimes Mg and Al, see e.g. \citealt{gratton}), the mechanism responsible for these variations has not yet been determined. A number of different candidates for this pollution has been proposed, namely AGB stars (e.g. \citealt{cottrell,denissenkov97,ventura2009,karakas2}), fast-rotating massive stars (FRMS, \citealt{decressin}), massive interacting binaries \citep{demink}, early disk accretion on low-mass protostars \citep{bastian}, or supermassive stars \citep{denissenkov}. All of these candidates can, with the right assumptions, explain parts of the observed behaviour, but each candidate struggles with its own set of problems, and no single polluter can explain the observed behaviour in full \citep{bastian2}.

A large part of the observational constraints on the polluter candidates comes from elemental abundance studies. Here, the presence of the Na-O anti-correlation, as well as the occasional Mg-Al anti-correlation, gives insight into the burning conditions that must have occurred in the polluters. The magnitude of the variations also gives clues to the amount of processed material that needs incorporation in the polluted population of stars, and the degree of dilution with pristine gas. However, even with the ever-increasing number of GC stars with accurately measured abundances, no consensus has been reached on the full explanation of the observed trends. Thus, it is necessary to try and provide additional diagnostics that can be used to constrain the nature of the polluters.

The measurement of Mg isotopes may provide additional pieces of information, supplementing what can be inferred from abundance ratios. In most stars the isotopic distribution is dominated by \mgi, which is mainly produced in type II supernovae, although some production can also happen in AGB stars \citep{wallerstein}. The amount of the two heavy isotopes, \mgii\ and \mgiii, is typically significantly less, with the solar value being 79:10:11, given as percentage \mgi :\mgii :\mgiii. The production of heavy Mg isotopes is rather complex, and can occur in multiple sites. In the context of variations of the Mg and Al abundances, one possible mechanism is AGB stars undergoing hot bottom burning (HBB), which will alter the distribution of the Mg isotopes. This process is very sensitive to the conditions in the nuclear burning zone, more so than just the bulk abundances (see e.g. \citealt{karakas}). Thus, when possible to measure magnesium isotopes in GC stars, these measurements can be used to yield detailed information about the polluters, compared to what elemental abundances can deliver.

Obtaining spectra of a quality where the isotopic distribution can be measured is observationally challenging, because it requires very high resolution spectra ($>80\,000$), combined with exquisite signal-to-noise ($\geq150$). Such measurements only exist for a few GCs, starting with the pioneering work of \citet{shetrone}, who used the observed variations in heavy isotopes to rule out deep mixing as the cause of the abundance variations in the GC M13. This provided a strong argument for the case of abundance variations being present already in the gas that formed the polluted generation of stars, the now commonly accepted internal pollution scenario. Measurements of Mg isotopic ratios have also been carried out for NGC6752 \citep{yong6752}, M13 \citep{shetrone,yongm71}, M71 \citep{melendezm71,yongm71} and $\omega$Centauri \citep{dacosta}. 

In all cases, a correlation has been observed between the amount of heavy Mg isotopes and [Al/Fe], which becomes particularly evident when the [Al/Fe] enhancement reaches values $\geq0.5$ dex. In the context of AGB stars, production of heavy Mg isotopes requires adopting the maximum values consistent with nuclear physics for the $^{25}\mathrm{Mg}(p,\gamma)^{26}\mathrm{Al}$ reaction \citep{ventura2008,ventura2011hbb,venturaMgAl2011}. The models also predict that most of the initial \mgi\ is converted to \mgii. An experimental model by \citet{venturaMgAl2011}, where the proton capture on \mgii\ is increased by a factor of two, is required to explain the most extreme [Al/Fe] enrichments, and also reduces the final amount of \mgii. However, the models still predict a \mgii/\mgi\ ratio that is significantly higher than the \mgiii/\mgi\ ratio, which is in contrast to the observations. \citet{karakas2006} on the other hand, achieves \mgiii/\mgi\ ratios higher than \mgii/\mgi\ in their models, but their adopted reaction rates also result in a net production of Mg, which is in contrast to the Mg-Al anti-correlation seen in a number of GCs. A similar behaviour is also observed for the most massive models in the computations of \citet{fishlock}.

The FRMS are also able to process some Mg, which results in a production of Al, at the expense of \mgi. However, this requires that the nuclear reaction rates for proton capture on \mgi\ are artificially increased by three orders of magnitude \citep{decressin}. This modification is also required to reproduce the Mg isotopic distribution in the case of NGC6752, which \citet{decressin} use for model comparison. Such drastic changes are well outside the total range of reaction rates determined from nuclear physics, which spans less than one order of magnitude, for the relevant temperatures \citep{Longland20101}. Using standard reaction rates, the FRMS provide a net production of Mg, mainly in the form of \mgi, so this would also result in a modification of the isotopic ratios. However, from the models one would expect a correlation between Mg and Al, which again contradicts the observations.

It has been shown by \citet{prantzos2} that the conditions required to activate the Mg-Al burning chain can be reached in the center of massive stars, but only at the very end of their main sequence evolution. There exist no known mechanism for transporting the processed material to the surface of the star, and release it in a slow fashion so that it can be retained within the cluster at this stage of the stellar evolution. Rather, at this point, the mass loss is dominated by a fast wind, which a typical GC will have difficult to retain \citep{decressin}. Finally we note that also the supermassive stars \citep{denissenkov} reach the appropriate conditions for activating the Mg-Al chain, but these models rely on some very particular conditions, and requires more exploration before they can be applied to GCs in full.

The GC 47 Tucanae is known to harbour at least two populations of stars, making itself known both through splittings of stellar sequences in the colour-magnitude diagram \citep{anderson,milone}, and through light element abundance variations (e.g. \citealt{cottrell,briley2,alves-brito,koch,cordero,thygesen}). Most pronounced is the anti-correlation between Na and O, but also a variation in [Al/Fe] is observed. Even though, due to its proximity, 47Tuc is exceptionally well-studied, a good understanding of its chemical evolution history is still lacking. \citet{ventura} proposed a pollution scenario based on AGB stars as the main source of pollution. They were able to explain a large part of the observed abundance variations, but as recently argued by \citet{bastian2}, none of the polluter candidates are able to explain the full range of observed abundance variations.

In this paper we add measurements of Mg isotopes in 13 stars to the chemical inventory of the cluster. This is a continuation of \citet{thygesen}, hereafter Paper 1, where we derived stellar parameters and abundances for a broad range of elements for the same sample of stars. By also adding measurements of the Mg isotopes, we can investigate the candidate polluters even further.

\section{Observations}
The observations for this project were obtained with the UVES spectrograph \citep{dekker} equipped with Image-Slicer \#3, mounted on the ESO VLT on Cerro Paranal, Chile. Each spectrum has a resolution of $R=110\,000$ and reached a signal-to-noise of $\sim$150 at the position of the MgH features, which is required for accurate determination of the distribution of magnesium isotopes. The targets were chosen from the cluster colour-magnitude diagram such that members of both the pristine and polluted population were included. We refer the reader to Paper 1 for further details about the observations and data reduction.

\section{Analysis}
The isotopic shifts of the atomic lines of Mg are much smaller than the natural broadening of the spectral lines, so in order to measure the isotopic mixture, one has to utilise molecular lines of magnesium hydride (MgH). We use the features from the electronic A-X transitions around 5135\,\AA, where the isotopic shifts are observed as an asymmetry in the red wing of the dominating \mghi\ feature. 

\subsection{Line selection}
\label{lineselection}
Traditionally, three molecular MgH features are used for the derivation of magnesium isotopes \citep{mcwilliam} at: 

\begin{itemize}
	\item 5134.6\,\AA, which is a blend of the $0-0Q_1(23)$ and $0-0R_2(11)$ electronic transitions. 
	\item 5138.7\,\AA, which is a blend of the $0-0Q_1(22)$ and $1-1Q_2(14)$ electronic transitions.
	\item 5140.2\,\AA, which is a blend of the $0-0R_1(10)$ and $1-1R_2(14)$ electronic transitions.
\end{itemize}

All of the MgH features mentioned above, suffer from blends with both atomic lines and molecular lines of C$_2$, CN and CH. However, it is believed that these three features are suffering the least from blends, compared to other MgH transitions in the vicinity. 

Compared to previous investigations of Mg isotopic ratios, we used new, updated line positions and level energies taken from \citet{shayesteh} for \mghi\ and from \citet{hinkle} for \mghii\ and \mghiii, where all previous studies have been using the pioneering work of \citet{bernath}. The changes in line position between our data and the work of \citet{bernath} were typically small, so this did not have a strong impact on the derived results, compared to employing the linelist used in earlier works. For CH we used line information from \citet{masseron}, whereas for C$_2$ we use \citet{brooke}, and \citet{snedencn} together with \citet{2014ApJS..210...23B} for CN. For the atomic blends, we used the line information from version3 of the Gaia-ESO survey linelist (Heiter et al. in preparation), with a few lines added from the VALD database \citep{vald}. We include atomic blends from Ti, V, Cr, Fe, Co, Y, Nb, Mo, Ce and Pr. We used the elemental abundances measured in Paper I where available. In cases where the elemental abundances could not be measured, we used scaled solar values.

We checked our line list against a spectrum of Arcturus, which is admittedly more metal-rich, and it was found that a few C$_2$ lines came out very strong in the synthesis, when Arcturus was fitted with the stellar parameters from \citet{ramirez}, using the C abundance from \citet{snedencn}. This points towards either problems with the \loggf\ values, the presence of 3D effects on selected C$_2$ lines, or mis-identified lines. This issue is particularly evident for the region around 5140\,\AA. In Arcturus, the MgH feature present here, cannot be fit using the C abundance from \citet{snedencn}, but requires additional depletion, even if the \mgii\ and \mgiii\ components are removed entirely. The overly strong C$_2$ lines in Arcturus in the 5140\,\AA\ region are so overestimated that a much better fit is obtained by simply removing them from the syntheses. Conversely,  we found that this is not advisable in the case of our stars. 

While the C$_2$ lines are too strong, at least for stars as metal-rich as Arcturus, they are also temperature-sensitive, increasing with strength with temperature, for the temperature range considered here. If the C$_2$ lines in question were left out when analysing our 47Tuc sample, this resulted in a spurious trend of the isotopic ratios with temperature when inspecting the individual results from the 5134.6\,\AA\ and 5140.2\,\AA\ features. These trends were not present when the strong C$_2$ lines were included in the line list. As such, the complete removal of the C$_2$ lines amounts to overcompensating. Rather, the culprit C$_2$ lines do appear to be present, but likely their \loggf\ values are too high. 

\begin{figure*}[htb!]%
\centering
\includegraphics[width=0.95\textwidth, trim = 2.5cm 2.5cm 0cm 0cm]{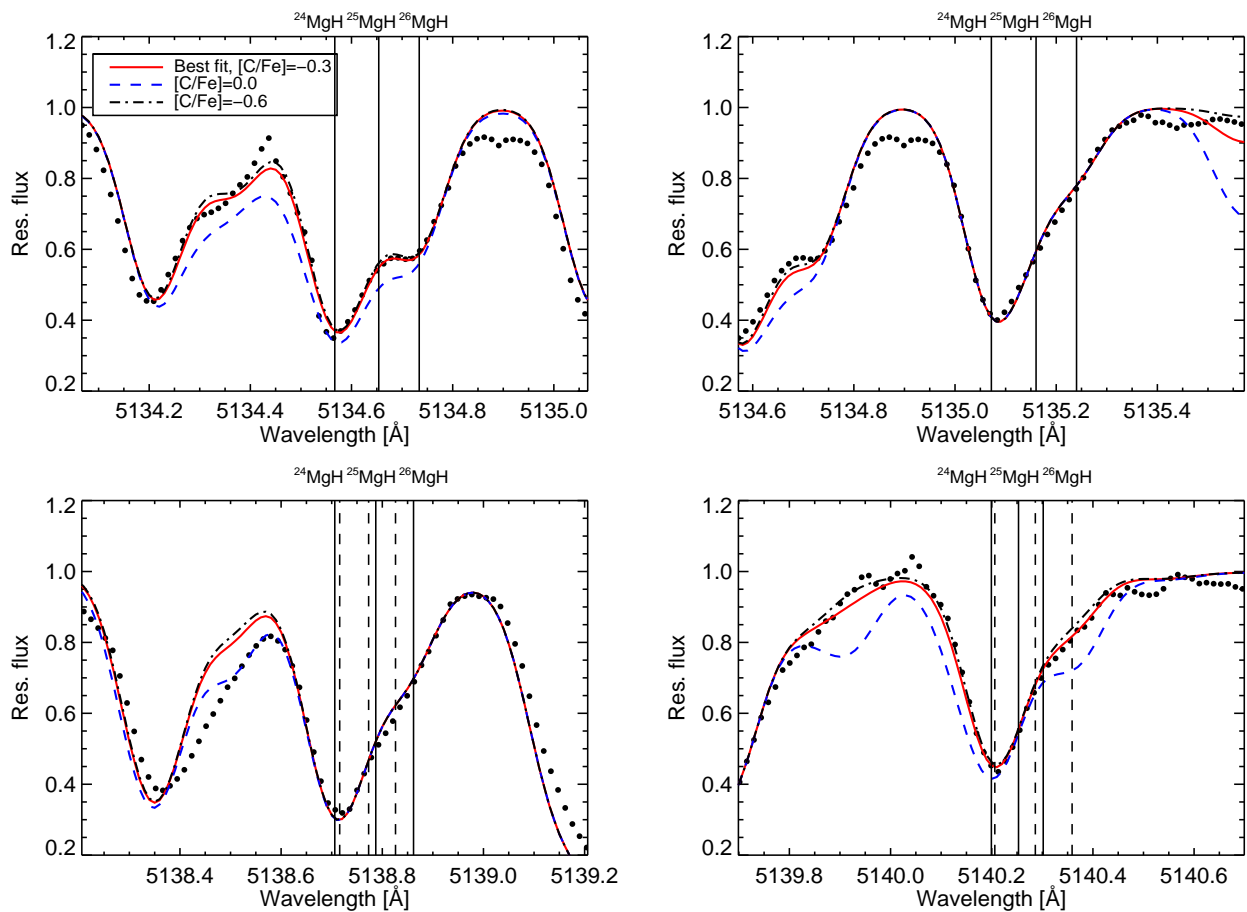}%
\caption{Synthesis of the four MgH features used to derive the isotopic fractions of Mg in star 5265. The red line shows the synthesis using best fitting [C/Fe] value, whereas the dashed lines show the change of the line shape when the [C/Fe] is changed by $\pm0.3$ dex around this value.}%
\label{varycarbon}%
\end{figure*}

From this exercise, we also identified a previously unused MgH molecular feature at 5135.1\,\AA\, that is also not sensitive to the C abundances (Fig.~\ref{varycarbon}). If we exclude the 5140\,\AA\ feature for Arcturus, we find the percentage distribution of  \mgi:\mgii:\mgiii\ = 82.3:9.0:8.7, if taken as a straight average. If we instead do a weighted mean, like adopted for the 47Tuc stars (see later) we find  82.0:9.8:8.2. Finally, if we include the 5140\,\AA\ feature (but deplete C by an additional 0.1 dex, compared to the value quoted in \citealt{snedencn}), we arrive at 83.6:9.2:7.2 for the distribution of the isotopes. This is in good agreement with the values found by \citet{mcwilliam}, who found 82:9:9; and \citet{hinkle} who reported 80:10:10. Keeping in mind the differences in line lists and analysis methods, we consider this agreement satisfactory.

In order to use the 5134.6\,\AA\ and 5140.2\,\AA\ features to derive Mg isotopic ratios, information about the C abundance is required. \citet{melendezm71} and \citet{yong6752} use two different regions to estimate the C abundance in their sample of giants, namely a C$_2$ feature at 5136\,\AA, and the C$_2$ features around 5635\,\AA. In our sample of stars, we cannot derive an useful C measure from the 5136\,\AA\ feature, but we were able to use the 5635\,\AA\ region, although the C$_2$ features were very weak. While this region shows multiple blends from other molecular species, it can be used to give an indication of the C abundance in our stars. The lines are already very weak, and are not sensitive to strong C depletion. As such, our derived values should only be taken as upper limits on the C abundance. As can be seen in Fig.~\ref{varycarbon}, the 5134.6\,\AA\ and 5140.2\,\AA\ MgH features are very sensitive to the C abundance, but nonetheless, for our 47Tuc stars, we are able to fit these two features with the C abundance estimated from the C$_2$ transitions in the 5635\,\AA\ region. 

However, keeping in mind the result from the test on Arcturus, even with a correct C abundance, the isotopic ratios derived from these two features should not be considered as reliable as what can be derived from the 5135.1\,\AA\ and 5138.7\,\AA\ features, which do not show such sensitivity. In particular the measurements from the 5140.2\,\AA\ feature will likely underestimate the true amount of the heavy isotopes.

When calculating the average Mg isotopic fractions, we only give half weight to the results from the 5134.6\,\AA\ and 5140.2\,\AA\ features, although we were able to fit the features with similar isotopic fractions to the 5135.1\,\AA\ and 5138.7\,\AA\ features in most cases. This was done for all stars in the sample, due to the likely \loggf\ problems discussed above. Full weight was given to the 5135.1\,\AA\ and 5138.7\,\AA\ features.  

\begin{figure*}[htb!]%
\centering
\includegraphics[width=0.95\textwidth, trim = 2.5cm 2.5cm 0cm 0cm]{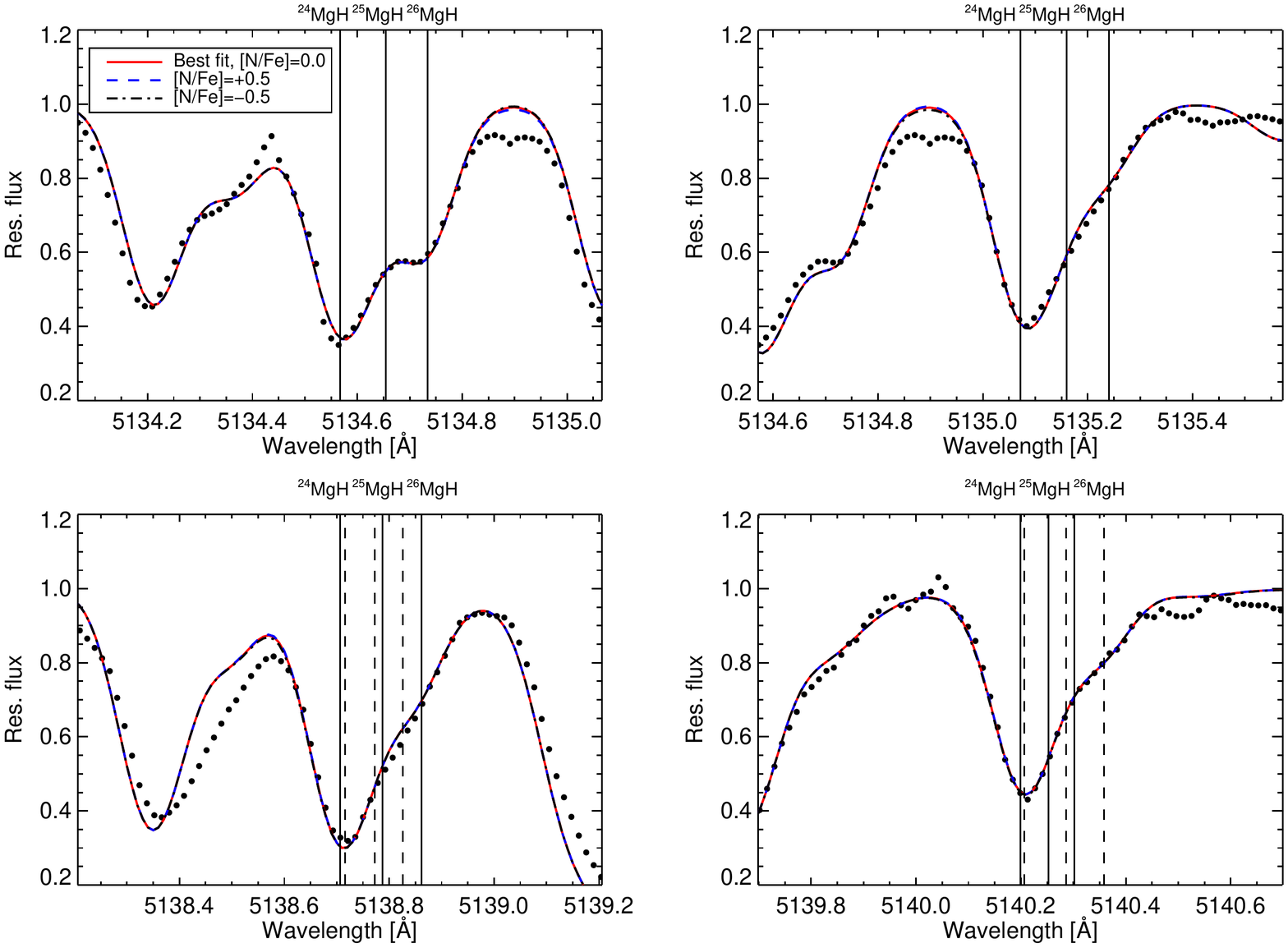}%
\caption{Synthesis of the four MgH features used to derive the isotopic fractions of Mg in star 5265. The red line shows the synthesis using the scaled solar [N/Fe], whereas the dashed lines show the change of the line shape when the [N/Fe] is changed by $\pm0.5$ dex around this value.}%
\label{varynitrogen}%
\end{figure*}

Since also the abundance of nitrogen is known to exhibit strong variations in GCs, this may affect the shape of the MgH regions as well, as blends of CN are also included in our line list. However, as illustrated in Fig.~\ref{varynitrogen}, even a 1 dex variation of N does not influence the line shapes. Thus, we do not consider the unknown nitrogen abundance to be an issue for our analysis.

In principle, also blends of TiO could influence our results. We tested this by calculating a synthesis for all four features, using all known transitions of TiO from \citet{plez}. We included lines of all stable Ti isotopes and used the solar isotopic mixture. This was done for our coolest star (star 5265), where the formation of TiO molecules should be most pronounced. Including the TiO features had negligible impact on the line shapes, and we discarded them in the synthesis for all remaining stars.

\subsection{Mg isotopes with MOOG}

We synthesised the MgH features using the 2013 version of the MOOG spectral synthesis code \citep{sneden1,sobeck,sneden2}, for our 1D analysis. We used the same ATLAS12 models as for the abundance analysis in Paper 1 as the basis for our syntheses. Initially a 20\,\AA\ piece of the spectrum was synthesised in order to set the continuum level. Using this normalisation, we determined the initial best fitting isotopic mixture by eye, where we varied both the isotopic mixture and the abundance of Mg, until a satisfactory fit was achieved. This was done individually for each of the considered regions. The isotopic mixture variations were accomplished by adjusting the \loggf-value of the molecular transition so that the fractional strength of each isotopic component corresponded to the fractional abundance of that isotope. With the quality of the observations available here, variations as small as 3\%\ for \mgii\ and \mgiii\ can easily be distinguished by eye. The abundance of Mg is a free parameter in the fits, and is solely used to adjust the overall strength of the MgH features. Following the procedure in \citet{melendezm71}, we revisited our broadening, by synthesizing Fe I lines at 6056.0, 6078.5, 6096.7, 6120.2 and 6151.6\,\AA. The broadening derived from these lines were subsequently adopted, assuming a Gaussian broadening profile for each star, representing the instrumental resolution, macroturbulence and rotational broadening as a whole. 

\begin{figure}%
\centering
\includegraphics[width=\columnwidth, trim= 2cm 3.5cm 2cm 3.5cm]{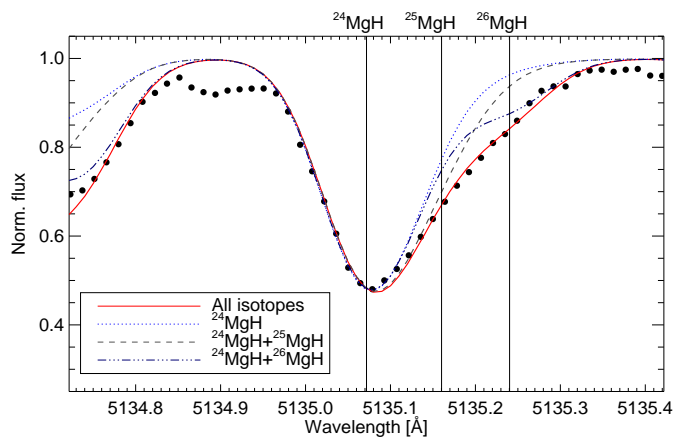}%
\caption{1D syntheses of the 5135.07\,\AA\ MgH feature for different isotopic components. The central wavelength of each isotopic transition is indicated with vertical lines.}%
\label{mgh-components}%
\end{figure}

Even though the isotopic splittings of the MgH features are small, it is immediately evident that all three isotopic components are required to produce a satisfactory match between the observed spectra and the syntheses. This is illustrated in Fig.~\ref{mgh-components} where we show the best-fitting syntheses for the 5135.07\,\AA\ feature in star 5265, including various isotopic components of the MgH feature. In all cases are \mgi+\mgii+\mgiii\ kept constant. Only the synthesis with all three components provide a satisfactory fit to the observed spectrum.

\begin{figure*}%
\centering
\subfloat[MgH 5134\,\AA]{%
  \includegraphics[trim= 10cm 6.5cm 0cm 0cm, width=0.45\textwidth]{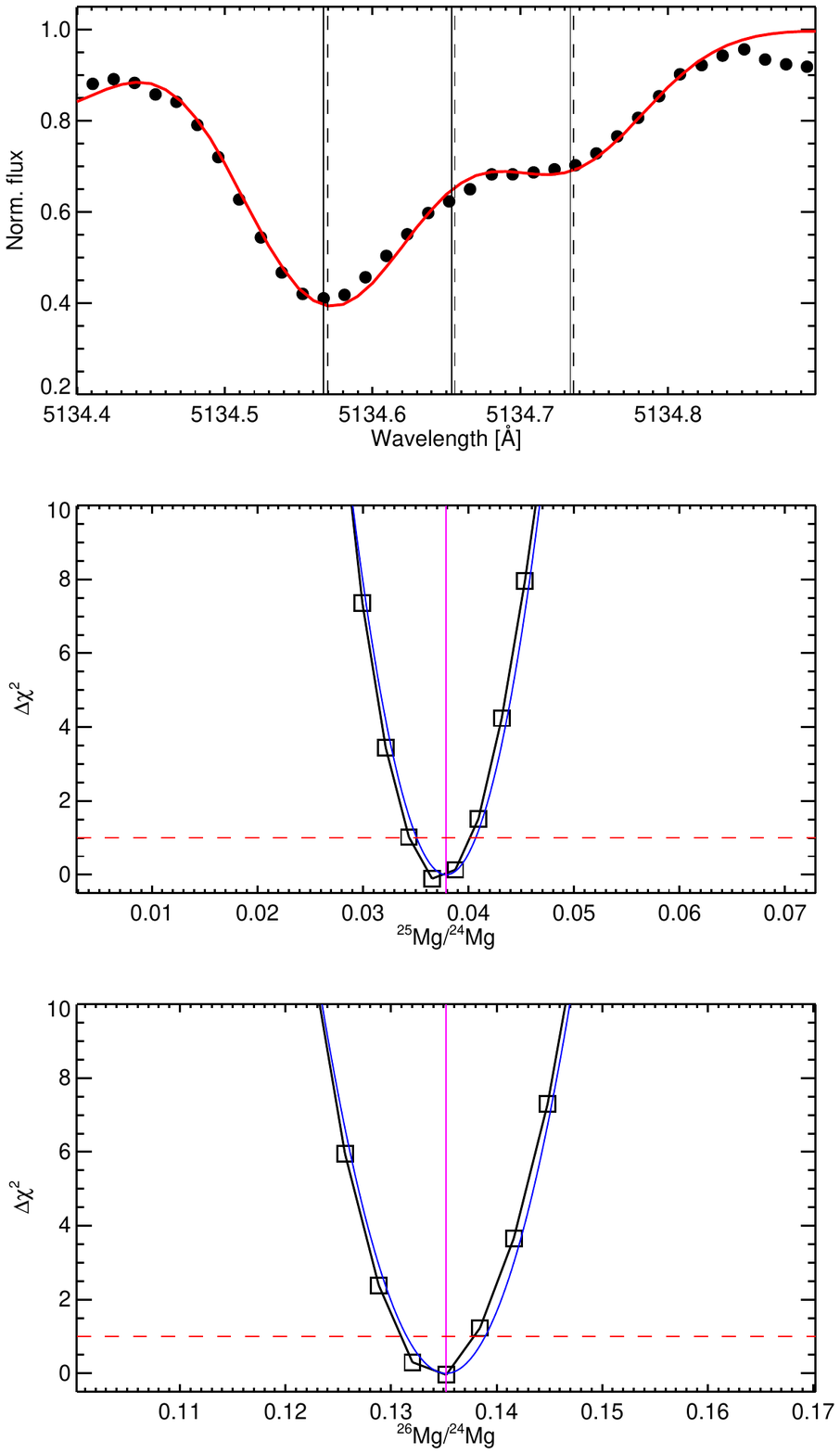}}
\quad
\subfloat[MgH 5135\,\AA]{%
  \includegraphics[trim= 10cm 6.5cm 0cm 0cm, width=0.45\textwidth]{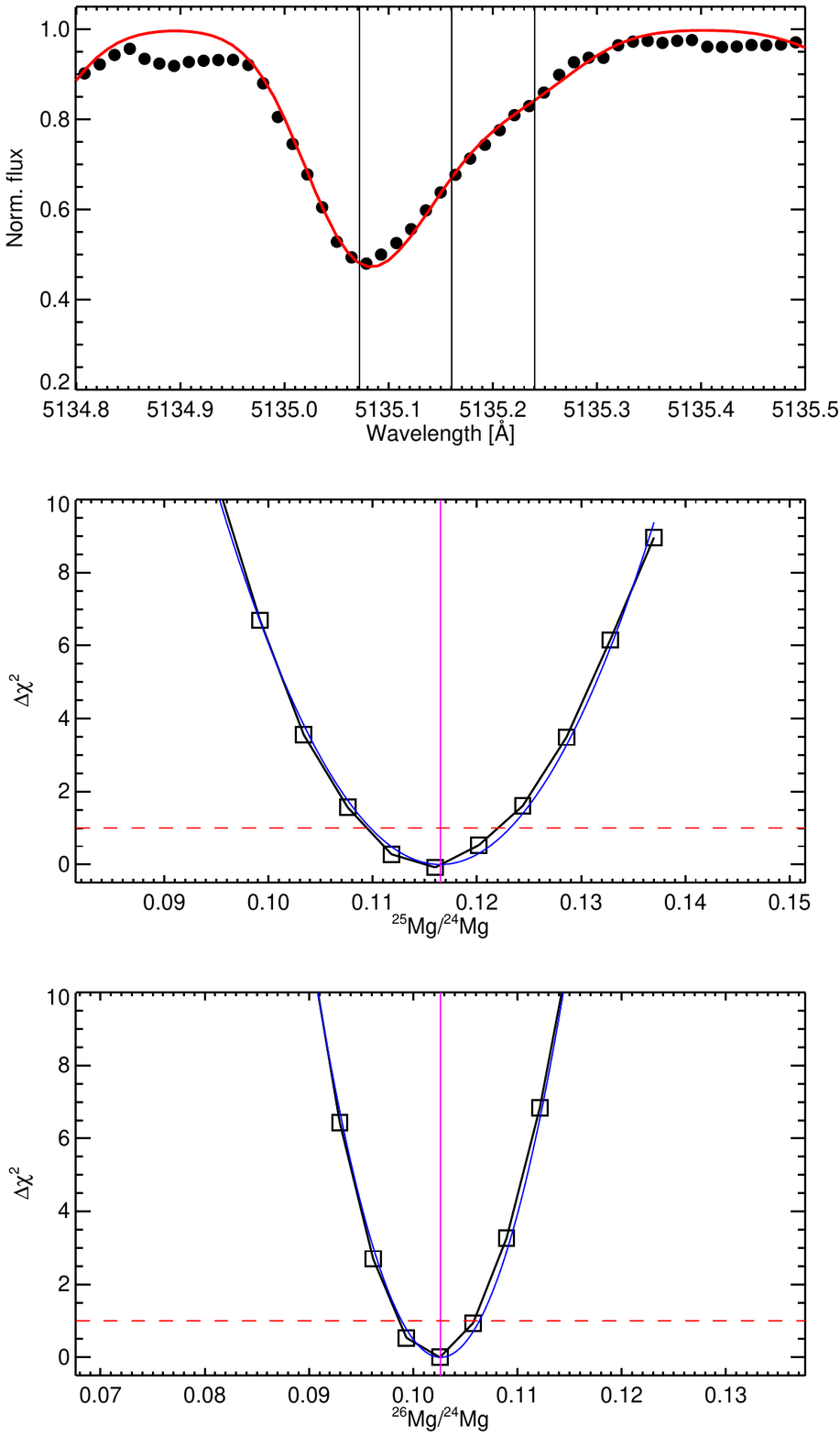}}
\caption{Line fits and $\chi^2$ distributions for the fitted isotopic ratios for the 5124\,\AA\ and 5135\,\AA\ MgH features in star 4794. The positions of the MgH transitions are indicated in the top plots. Vertical, magenta lines show best-fitting values in the $\chi^2$ plots.. The red dashed lines indicate the $1\sigma$ fitting precision.}%
\label{chisquared}%
\end{figure*}

\begin{figure*}%
\centering
\subfloat[MgH 5138\,\AA]{%
  \includegraphics[trim= 10cm 6.5cm 0cm 0cm, width=0.45\textwidth]{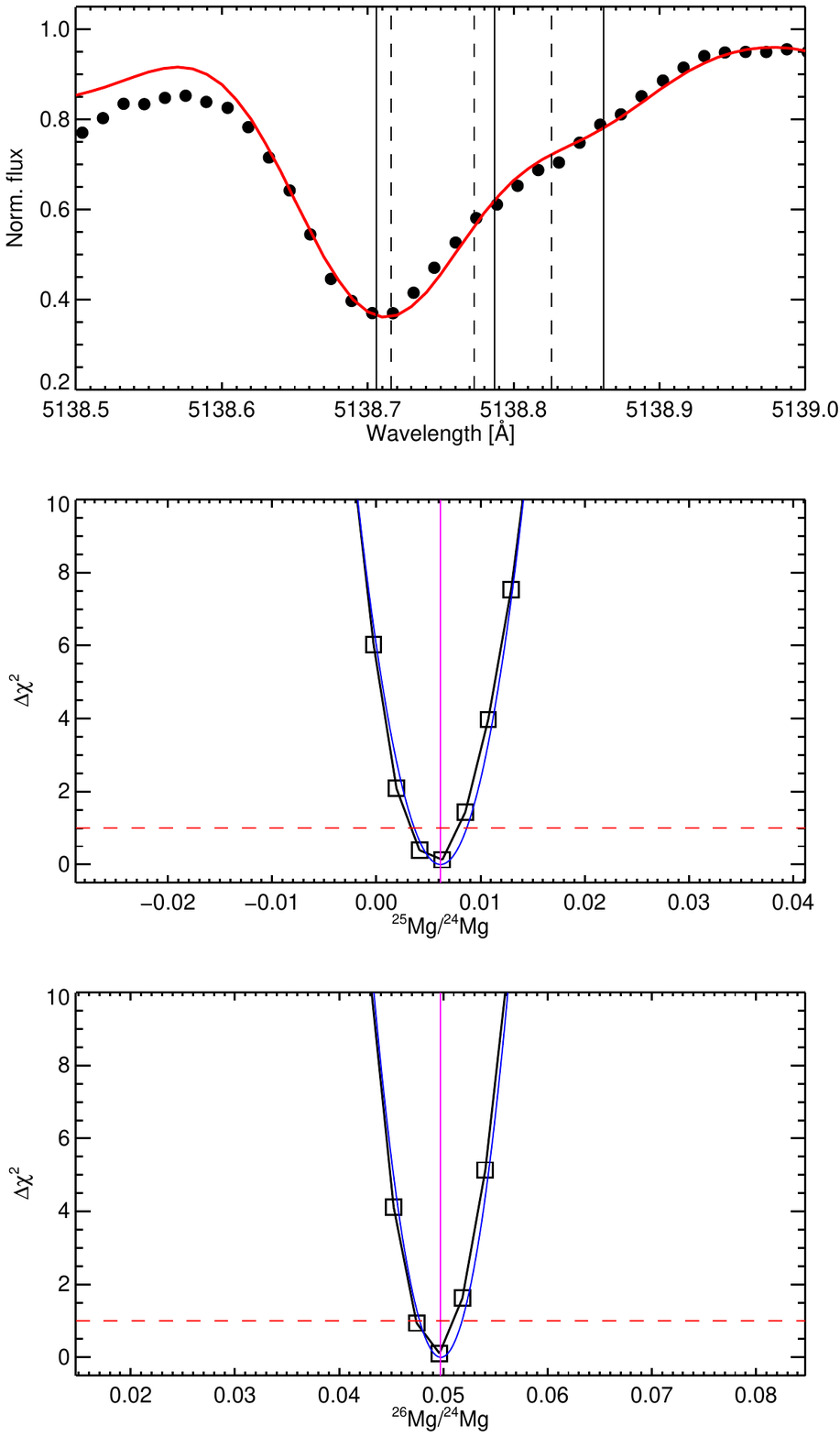}}
\quad
\subfloat[MgH 5140\,\AA]{%
  \includegraphics[trim= 10cm 6.5cm 0cm 0cm, width=0.45\textwidth]{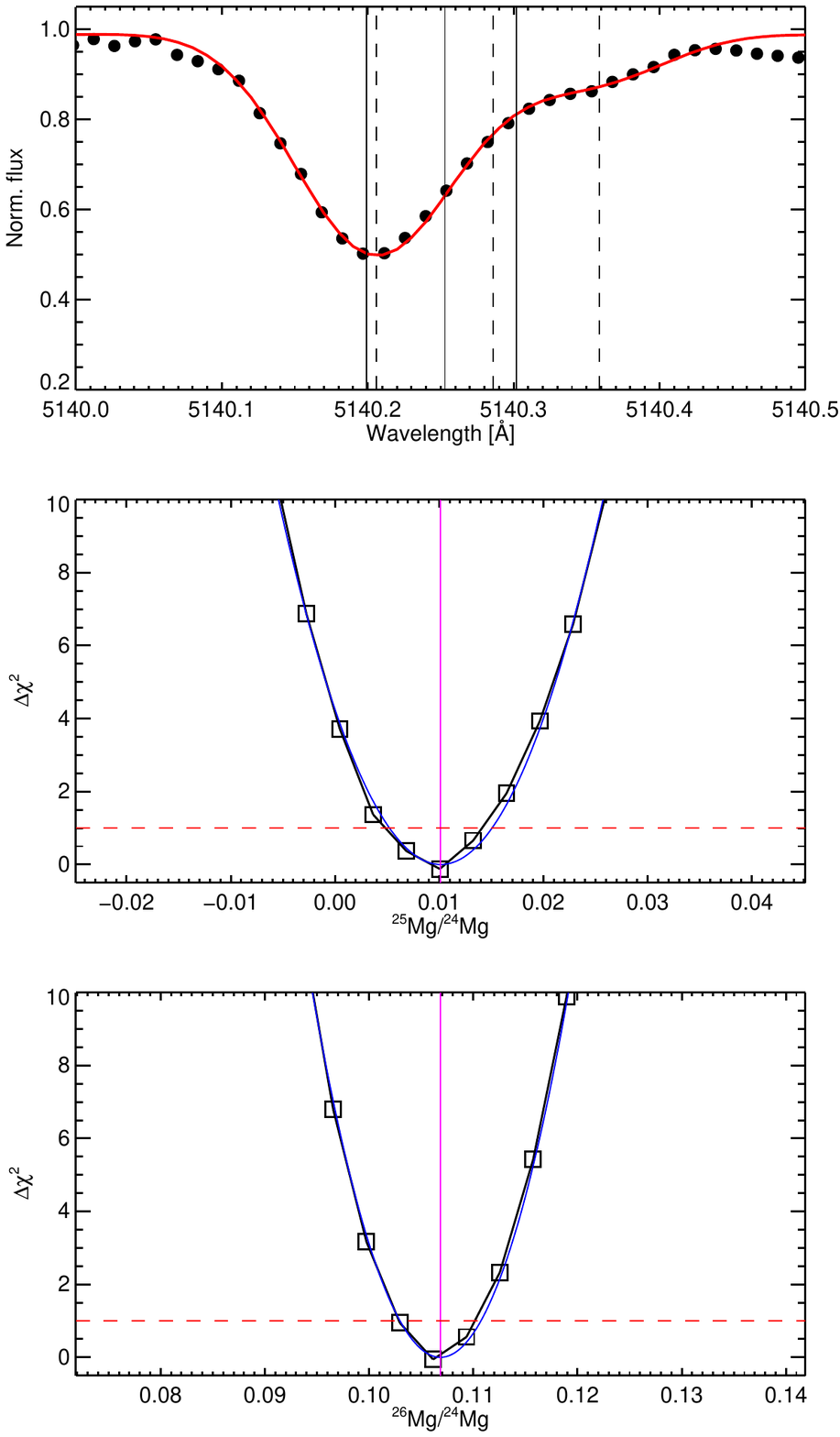}}
\caption{As Fig.~\ref{chisquared}, but for the 5138\,\AA\ and 5140\,\AA\ features.}%
\label{chisquared2}%
\end{figure*}

Using our by-eye fits as a first estimate of the isotopic ratios, we refined the synthesis, determining the best fit by doing a $\chi^2$-minimisation between the observed spectrum and a grid of synthetic spectra following the method of \citet{yong6752}. This was done iteratively, in each step varying the isotopic ratios $^{25,26}$Mg$/$\mgi\ and total magnesium abundance. Initially, we explored a large parameter space, but for the final result we adopted a finer sampling to refine our results. Each of the four MgH features used was fitted independently. As an example, we present the best-fitting syntheses, as well as curves showing $\Delta\chi^2=\chi^2-\chi^2_{min}$ for all four features in the star 4794 in Figs.~\ref{chisquared} and \ref{chisquared2}. Here, we take $\Delta\chi^2=1$ as the $1\sigma$ fitting precision. It is evident that the fitted ratios are detections at least at the $3\sigma$ level, for both the \mgii/\mgi\ and \mgiii/\mgi\ ratios. 

In total, more than 1300 different syntheses were calculated in each iteration for each feature. The differences between the isotopic fractions found by eye, compared to the $\chi^2$-fitting were small, typically 1\% but never higher than 3\%. The final isotopic mixture from this method is given as a weighted mean of the individual results, with full weight given to the 5135\,\AA\ and 5138\,\AA\ features. We only gave half weight to the results from the features at 5134\,\AA\ and 5140\,\AA, which suffered from C$_2$ molecular blends.

\subsection{Mg isotopes with {\sc CO5BOLD}/Linfor3D}
When deriving the Mg isotopic ratios from the line asymmetries in the MgH features it is important to take into account all potential effects that could create such asymmetries. Blending with other atomic and molecular lines is one effect that could cause discrepancies in the observed isotopic ratios, as discussed above. Also the convective motions of gas in the stellar atmospheres are known to cause line asymmetries (see e.g. \citealt{dravins}). Convection is an intrinsically multi-dimensional phenomenon and cannot be modelled in 1D. In the ATLAS12 models, convection is approximated by the mixing-length formulation, and the usual macro- and micro-turbulence quantities are used to parametrise large- and small-scale gas motions respectively.

\subsubsection{Model setup}

Three-dimensional, hydrodynamical atmospheres do not rely on the mixing-length theory (MLT) approximation or similar approximate descriptions for treating convective energy transport, nor on micro-/macro-turbulence parameters for the modelling of convective broadening of spectral lines. We used the {\sc CO5BOLD} atmospheric code \citep{co5bold} to calculate two box-in-a-star 3D LTE models. We calculated models with parameters representative for the less evolved stars in our sample. The model parameters are listed in Table~\ref{model3d}. In 3D atmosphere models, \teff\ is not a control parameter as in the 1D case. Rather, it is an outcome of the simulation, indirectly controlled by the inflowing entropy. Hence why the 3D model \teff\ is not the same in the two cases. We did not calculate 3D models for our more evolved giants, since the horizontal temperature fluctuations become so large in these models that the radiative transfer code breaks down and provides unphysical results. In addition, the box-in-a-star approximation is challenged for low gravity models. Rectifying these problems would be a major theoretical undertaking, well beyond the scope of this paper.

\begin{table}%
\centering
\caption{Model parameters for our {\sc CO5BOLD} 3D models.}
\begin{tabular}{ccccccc}
\hline
ID & \teff & \logg & [Fe/H] & [$\alpha$/Fe] & model time\\
\hline\hline
HiMet & 3970K & 1.50 & $-0.50$ & $+0.2$ & $1472$ h\\
LoMet & 4040K & 1.50 & $-1.00$ & $+0.4$ & $1111$ h\\
\hline
\end{tabular}
\label{model3d}
\end{table}

\begin{figure*}%
\centering
\includegraphics[width=\columnwidth, trim = 2cm 3.5cm 2cm 3.5cm]{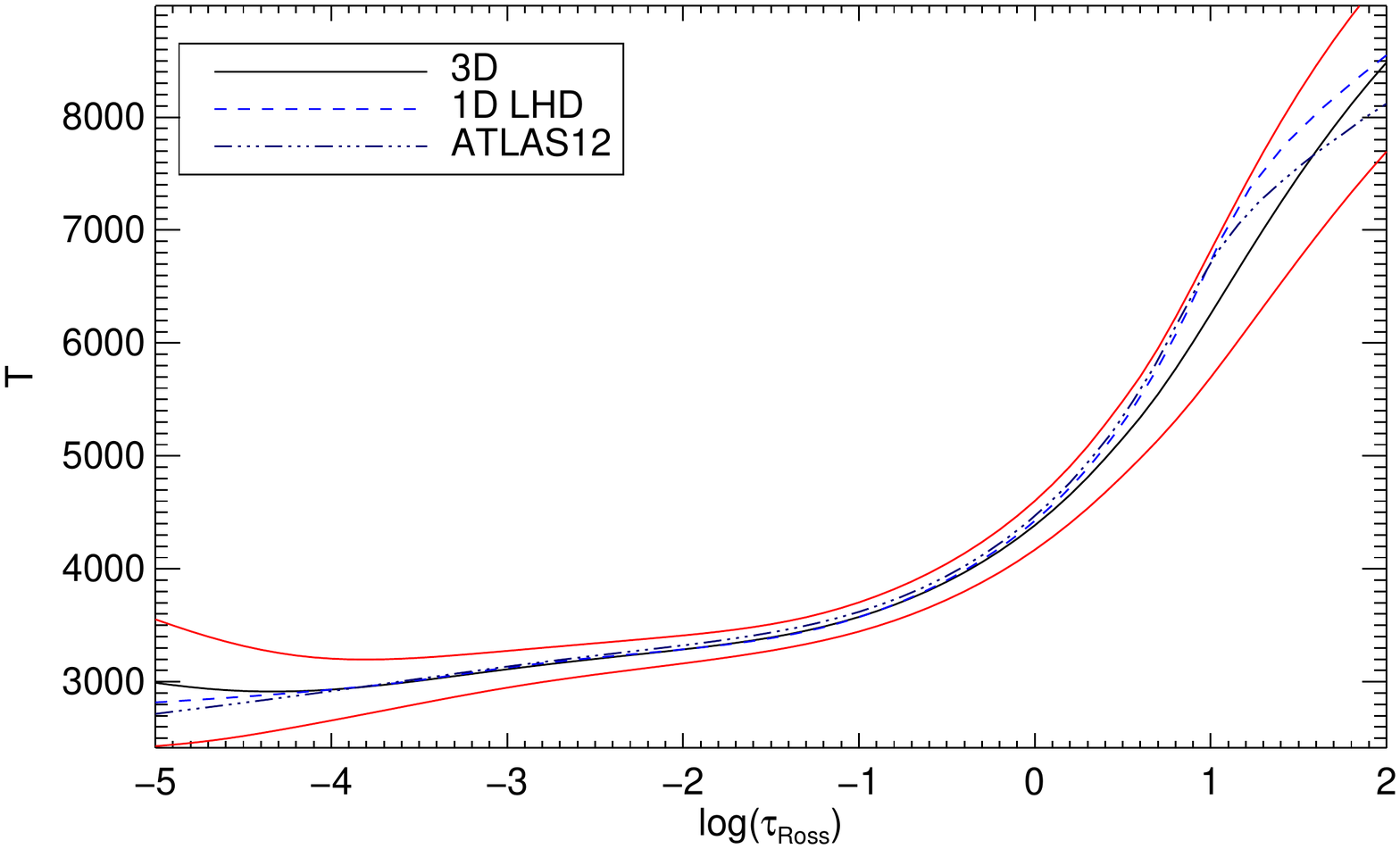}%
\includegraphics[width=\columnwidth, trim = 2cm 3.5cm 2cm 3.5cm]{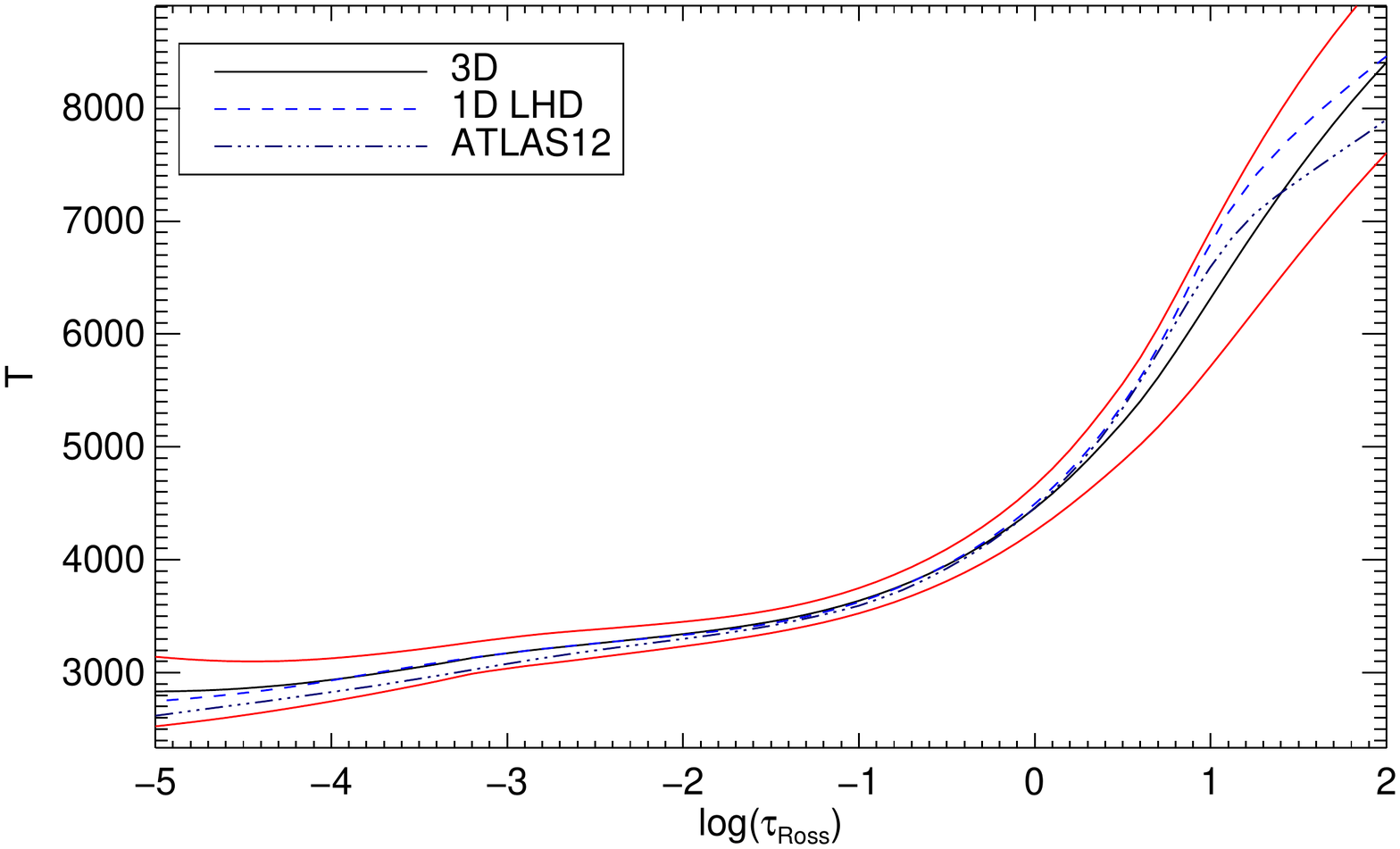}
\caption{Average thermal structure of the 3D models (solid line), the equivalent 1D ATLAS12 (triple dot-dash), and LHD (dash) models. The red lines indicate the RMS variation of the temperature in the 3D models. Left panel: HiMet, right: LoMet.}%
\label{struct3d}%
\end{figure*}

One of the main results of 3D hydrodynamical modelling of stellar atmospheres, is the prediction of average atmospheric temperature stratifications that are in general significantly different from the ones predicted by traditional 1D models. This is particularly evident for the lowest metallicities (e.g. \citealt{collet}), where a net cooling effect is manifested in 3D. In Fig.~\ref{struct3d} we plot the average thermal structure of the two 3D models together with the structure of the ATLAS12 models ($\alpha_\mathrm{MLT}=1.25$), and two reference LHD \citep{caffau2007} 1D models ($\alpha_\mathrm{MLT}=0.5$), which are calculated with the same input physics as the {\sc CO5BOLD} models. Also indicated are the mean temperature fluctuations in our 3D models. As is evident, at the metallicities treated here, the average thermal structure of the photospheres in the 3D models are nearly identical to the equivalent 1D models. On the other hand, the convection still causes significant variation in the thermal structure of our 3D model. The similarity of the average structures implies that differences between the 1D and 3D synthesis are mostly related to the temperature fluctuations, rather than changes in the overall atmospheric structure. 

\subsubsection{3D synthesis}
\begin{figure}%
\centering
\includegraphics[width=\columnwidth]{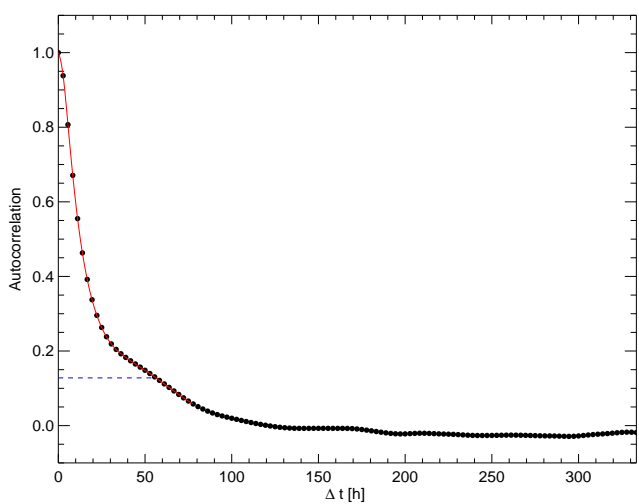}%
\caption{Auto-correlation of the horizontal grey intensity for the HiMet model. The blue, dashed line, shows the value for the time step between two consecutive snapshots.}%
\label{autocorr}%
\end{figure}

With the models computed, we selected 20 model snapshots, with near-equidistant time steps as a basis for the spectral synthesis. These were selected such that they covered nearly the full model time, to give good temporal coverage. Since the convective motions are responsible for a large part of the line asymmetries, the snapshots should ideally be sampled so that they represent uncorrelated convective patterns in the simulation. By computing the auto-correlation between the horizontal, gray intensities for all our model instances, we can gain insight into the typical lifetime of a convective cell. Having computed the auto-correlation, we performed a spline interpolation for a range of model times, so that our snapshot time step was included. From this, the value of the auto-correlation between two consecutive time steps could readily be determined. This showed that there is typically a 13\% correlation between our individual snapshots (Fig.~\ref{autocorr}), which is sufficiently uncorrelated for our purpose. We note that the time axis is truncated, so that the decrease in the auto-correlation is visible. The total length of the simulation is significantly longer than indicated in the figure ($\sim1400$h).

Furthermore, we ensured that the mean \teff, and RMS scatter of the \teff\ in our snapshot selection was nearly identical to the \teff\ and RMS found from the complete model grid. This represents a compromise between computational time and model coverage, so that the temperature variations from the 3D model are kept intact in the spectral synthesis. For each snapshot we calculated the full 3D radiative transfer using the Linfor3D code\footnote{\url{http://www.aip.de/Members/msteffen/linfor3d}}, to yield the emergent spectrum. The snapshot syntheses were subsequently averaged and normalised, to yield the final, synthetic spectra.

For each 3D model, we calculated a grid of syntheses with $-0.50<\Delta$[Mg/H]$<0.50$, in steps of 0.2, relative to the input mixture. The isotopic fractions of the heavy Mg isotopes were varied between $0<\mathrm{frac}(^{25}$Mg) $<0.21$ and $0<\mathrm{frac}(^{26}$Mg) $<0.21$ in steps of 0.03. The isotopic fraction of $^{24}$Mg was calculated as \mgi $=1.0-$\mgii$-$\mgiii. The relative strength of the isotopic components of the MgH lines were set by adjusting their \loggf\ value. Similar to the 1D syntheses, the Mg abundance was used to adjust the MgH feature strengths. We used an abundance mixture identical to the mixture used in the model computation as a starting point of our syntheses, as stated in Table~\ref{model3d}. The line list is identical to what was used in MOOG.

\begin{figure}%
\centering
\includegraphics[width=\columnwidth,trim = 2cm 3.5cm 2cm 3.5cm]{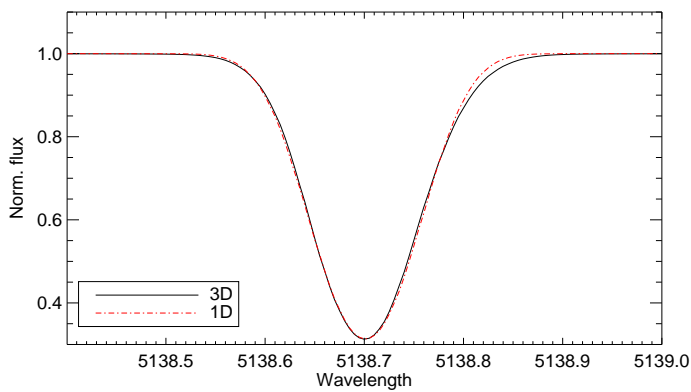}%
\caption{Comparison of the syntheses of the 5138\,\AA\ feature for \mghi. Note the line asymmetry in the red wing. The syntheses have been scaled to have the same overall strength.}%
\label{asymmetry}%
\end{figure}

Due to the velocity fields in the atmospheres of the 3D models, lines synthesised in 3D will exhibit a small velocity shift, relative to an equivalent 1D synthesis. To account for this, we synthesised a single $^{24}$MgH line in both 1D and 3D and cross-correlated the two spectra to determine the velocity shift. This gave shifts of 0.26\,\kms\ and 0.38\,\kms\ for the [Fe/H] $=-0.50$ and $-1.00$ models, respectively. By synthesizing only a single line, we bypass potentially different behaviour of line-blends between 1D and 3D, which could mimic an overall velocity shift. Before performing any analysis, we shifted the 3D spectra by this amount, to be as consistent with the 1D analysis as possible. It was, however, found that some additional velocity shifts were needed for some features, which was also the case for our 1D fits. The synthesis of the single $^{24}$MgH transition also revealed that 3D effects introduced a weak line asymmetry (Fig.~\ref{asymmetry}), which is expected also to be present for the two heavy MgH components of the feature. This can explain part of the different line shapes between 1D and 3D.

After shifting the spectra, the final 3D spectrum was convolved with the instrumental profile of UVES to yield the broadened spectrum. We assumed a Gaussian profile. In the absence of strong rotation, this should account for all the line-broadening present in the observed spectra, because the effects of micro- and macroturbulence arises as a natural consequence of the gas motions in the 3D models. Because this work deals with cool, evolved giants, we consider them non-rotating for all practical purposes, and did not impose any additional broadening to the spectra. This is further justified by the similarity of the V$_{\mathrm{macro}}$ measurements reported in Paper 1. With the syntheses calculated for both models, we performed a simple, linear interpolation in metallicity between the spectra, to match the observed metallicities of our stars. The grid of syntheses were subsequently given as input to the \texttt{Fitprofile} code, described below, to fit the observed spectra. 

Due to the issues with the C$_2$ blends discussed earlier, we chose only to calculate 3D synthesis for the 5135\,\AA\ and 5138\,\AA\ features, which do not suffer from strong carbon blends. If we were to synthesise also the 5134\,\AA\ features and 5140\,\AA\ features, it would require adding an extra dimension to our parameter space for the syntheses, namely C abundance variations. Since we are already varying the Mg abundance, and the fractions of $^{24,25,26}$Mg, it would become very computationally expensive to also include variations in the carbon abundance.

\begin{figure*}%
\centering
\includegraphics[width=\columnwidth, trim = 2cm 3.5cm 2cm 3.5cm]{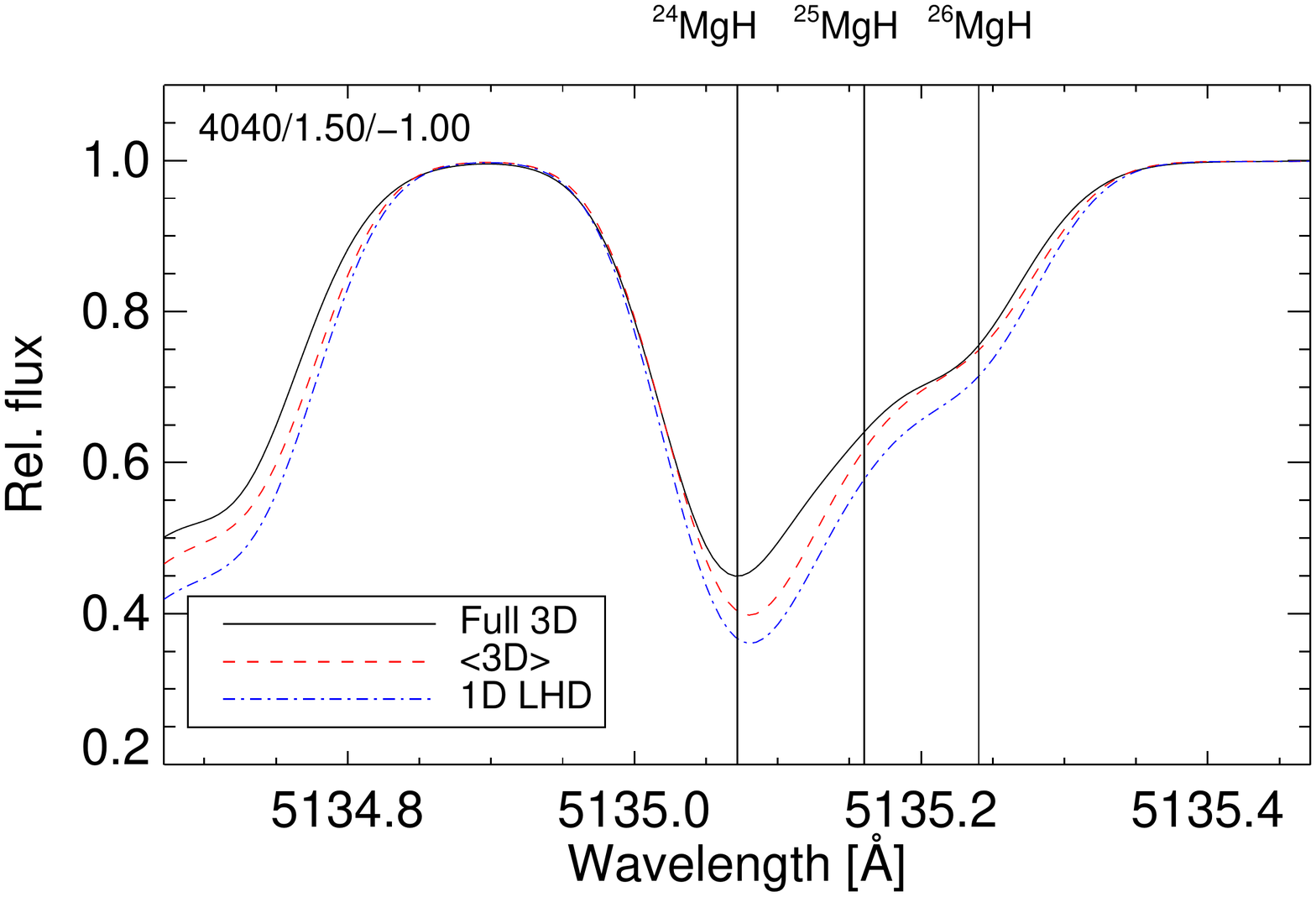}%
\includegraphics[width=\columnwidth, trim = 2cm 3.5cm 2cm 3.5cm]{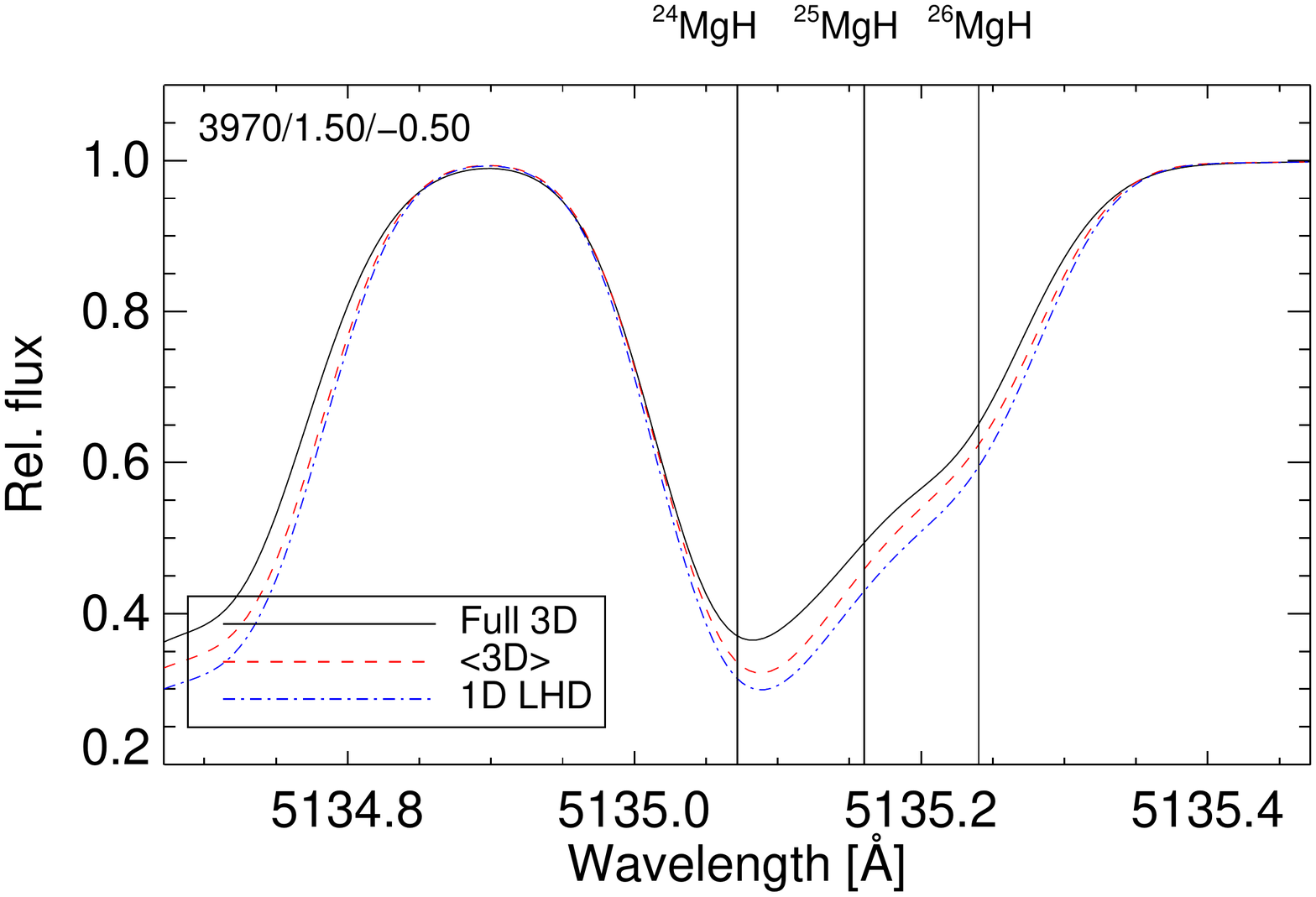}%
\caption{Syntheses of the MgH feature at 5135.07\,\AA. Shown is the full 3D synthesis (solid), the <3D> synthesis (dashed) and the 1D LHD synthesis (dot-dashed). Indicated is also the central position of each of the MgH components. Left: LoMet, right: HiMet.}%
\label{lowmetsynth}%
\end{figure*}

\begin{figure*}%
\centering
\includegraphics[width=\columnwidth, trim = 2cm 3.5cm 2cm 3.5cm]{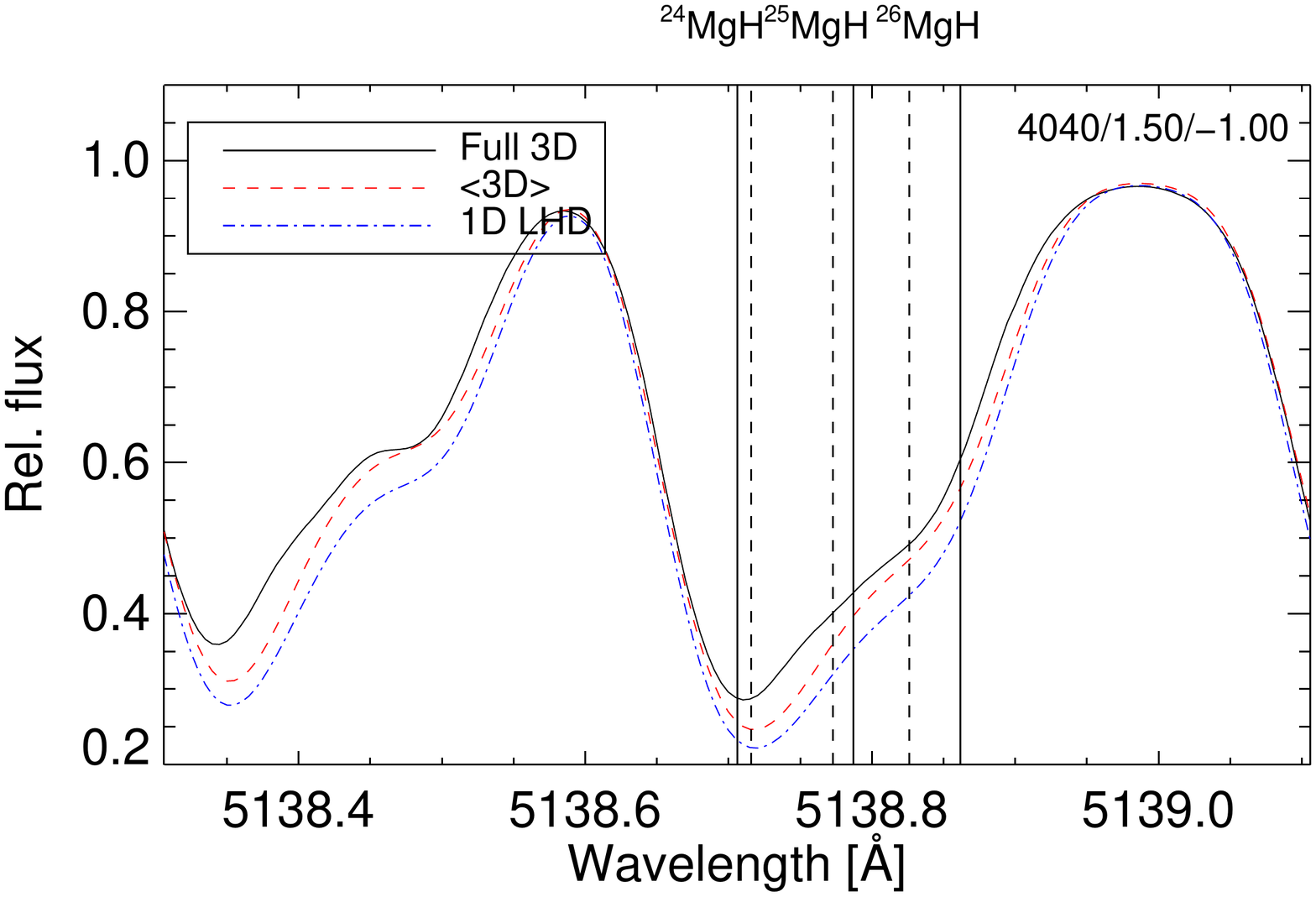}%
\includegraphics[width=\columnwidth, trim = 2cm 3.5cm 2cm 3.5cm]{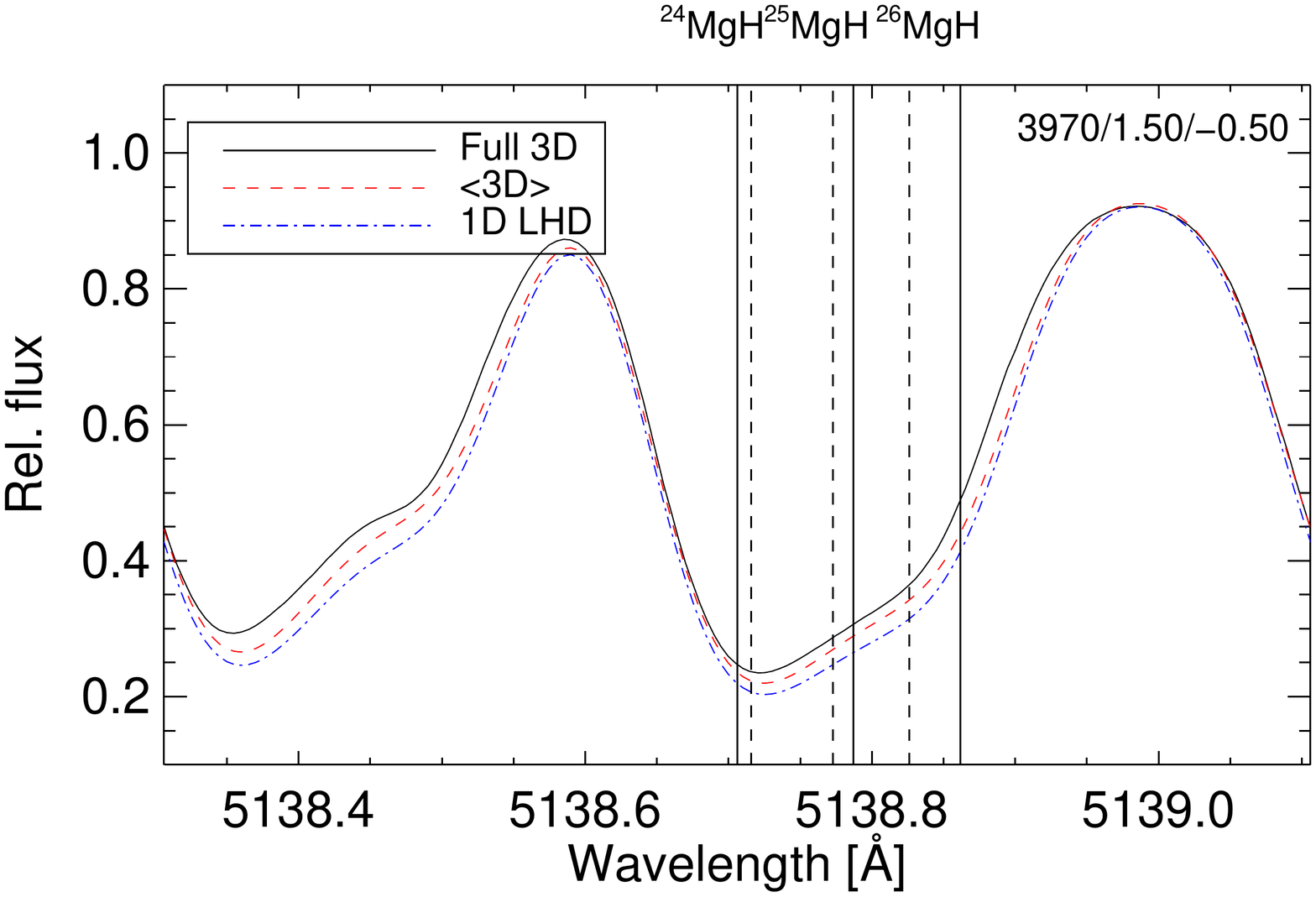}%
\caption{As Fig.~\ref{lowmetsynth}, but for the MgH feature at 5138.71\,\AA. This feature is a blend of two MgH transitions, indicated with solid and dashed vertical lines.}%
\label{highmetsynth}%
\end{figure*}

In Figs.~\ref{lowmetsynth}, \ref{highmetsynth} the effects of the 3D atmospheres are illustrated. Three different syntheses are shown. A full 3D synthesis, a synthesis using the horizontally averaged 3D temperature structure (<3D>), and finally a synthesis using a 1D atmosphere calculated with the same input physics as the 3D model. The HiMet and LoMet syntheses both have the same Mg-depletion (-0.10 dex) relative to the input mixture. As is evident, both the strength of the features, as well as the line shapes, change between 1D and 3D. Whereas some differences can be observed between 1D and <3D>, it is clear that in this case, one needs the full information from the 3D synthesis to capture the differences between 1D and 3D. This underlines that the main reason for the different behaviour of the MgH features is related to the temperature fluctuations and not changes in the average thermal structure. Interestingly, \citet{ramirez2} found that the overall strength of the MgH features were increasing when 3D atmospheres were applied to a K-dwarf model, where we, on the other hand find that the features are decreasing slightly in strength, relative to a standard 1D synthesis.

\begin{figure*}%
\centering
\includegraphics[width=\columnwidth, trim=7cm 8cm 7cm 0cm]{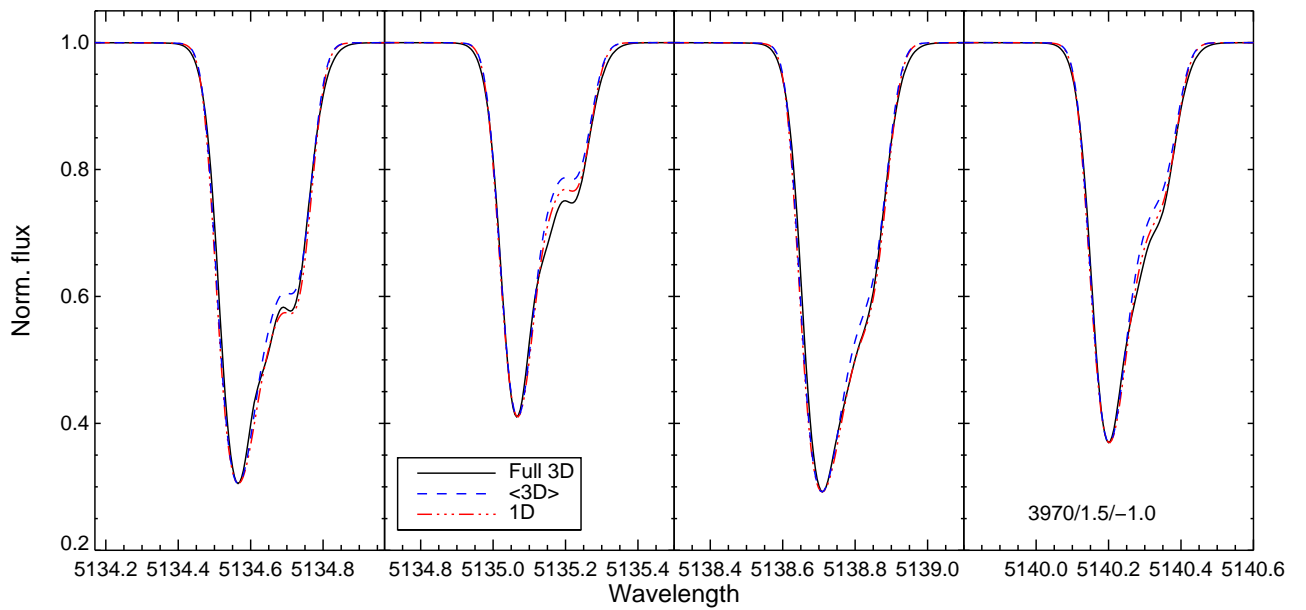}%
\caption{Syntheses of only the MgH transitions for all four features used. Solid line: Full 3D, dashed line: <3D>, and dot-dashed line: 1D. All syntheses have been normalised to same \mghi\ strength.}%
\label{mghonly}%
\end{figure*}

Naturally, both the MgH lines and the lines of the blending species will feel the effect of the 3D model atmospheres, and it is thus interesting to inspect whether the resulting changes in the line asymmetries are mainly due to effects on the MgH features themselves, or due to the blending species responding differently to 1D and 3D atmospheres. In Fig.~\ref{mghonly} we plot syntheses of all four MgH bands for our low-metallicity model, not including any blending lines. The syntheses have been normalised to have the same strength of the \mghi\ feature, and the 1D syntheses have been broadened to match the intrinsic broadening from the 3D syntheses. The normalisation to the same strength, allows for a better comparison of line asymmetries, but we note that without the normalisation, the MgH bands are weaker in 3D, compared to 1D, as also seen for the syntheses, including all blending species. It is clear that the effect of 3D on the MgH transitions themselves are rather small and most pronounced for the 5135\,\AA\ feature. Thus, this illustrates that the differences we find between 1D and 3D have a significant contribution from the blending species, at least for the 5135\,\AA\ and 5138\,\AA\ features. Whether this is the case also for the two remaining features cannot be judged from this exercise, and will require resolving the issue with the C$_2$ blending lines, as discussed earlier.

\subsubsection{Fitting the 3D syntheses with \texttt{Fitprofile}}
\input{fitprofile}

\subsection{Estimating uncertainties}
When one has spectra of such exquisite quality available as is the case here, the precision with which the individual features can be fit is remarkable. For our 1D analysis, we determined the minimum $\chi^2$-value for the three fitting parameters, \mgii/\mgi, \mgiii/\mgi\ and $\log\epsilon$(Mg). This was done for each of the four features individually. Typical fitting precision of the ratios are $\pm0.005$.

Whereas the individual fits are extremely precise, the feature-to-feature agreement is not as satisfactory. For most stars, the 5134\,\AA\ feature yields systematically higher \mgiii\ fractions, up to a factor two above the mean value. On the other hand, the 5135\,\AA\ feature yields \mgii\ fractions that are higher by a similar amount, relative to the typical mean. The former may be linked to the unknown carbon abundance, but we note that other studies of Mg isotopes find a similar behaviour for this line, even in cases where the carbon abundance is known. If carbon is enhanced slightly, this would result in a lower fraction of the heavy Mg isotopes. However, any enhancement of carbon would make it close to impossible to fit the 5140\,\AA\ feature, so this is unlikely to be the only explanation for the disagreement. We consider unknown blends to be the most likely cause of the disagreements between the features. These inconsistencies mean that the uncertainty of our final isotopic ratios become dominated by the feature-to-feature scatter. For the final values, we report the weighted mean of the isotopic fractions determined from each of the four features, and calculate the uncertainty, from line-to-line scatter, as the weighted standard error on the mean, with half weight given to the 5134\,\AA\ and 5140\,\AA\ features as discussed earlier.

Uncertainties in the atmospheric parameters of the stars will also contribute to the uncertainties of the Mg isotopic fractions. To investigate this, we repeated the fits of each of the four features, using atmospheric models perturbed by the uncertainties of the stellar parameters, following the method in Paper I. We use star 6798 as a representative example again, as it has parameters in the middle of our range. We then assume that the influence of the parameter changes found for this star, is representative for our full sample. In Table~\ref{mguncertainty} we report the changes to the mean value of the isotopic fractions, when the atmospheric parameters are changed. For the total influence of atmospheric parameters on the uncertainty, we take the mean of the absolute change of the fractions, from each parameter perturbation, and add them in quadrature. We note here that we keep the continuum and the velocity shift fixed in all cases, and thus the real uncertainty will likely include terms also reflecting this, as well as influence of unidentified blends and uncertainties in the abundances of the blending atomic species. We assume that such effects are small, and do not include them in our uncertainty budget.

\begin{table}%
\centering
\caption{Change in mean value of the isotopic percentages of Mg due to atmospheric uncertainties, relative to the best fit. Shown here for star 6798.}
\begin{tabular}{lrrr}
\hline
$\Delta$Param. & $\Delta^{24}$Mg & $\Delta^{25}$Mg & $\Delta^{26}$Mg \\
\hline\hline
$\Delta$\teff $=+80$K     &  0.2 &  0.0 & $-0.2$ \\
$\Delta$\teff $=-80$K     & $-1.0$ &  0.2 &  0.8 \\
$\Delta$\logg $=+0.2$     &  0.0 & $-0.1$ &  0.1 \\
$\Delta$\logg $=-0.2$     & $-0.6$ &  0.4 &  0.2 \\
$\Delta$\vmic $=+0.1$\,\kms &  0.2 &  0.0 & $-0.2$ \\
$\Delta$\vmic $=-0.1$\,\kms & $-0.5$ &  0.3 &  0.3 \\
$\Delta$[M/H] $=+0.15$    & $-1.1$ &  0.7 &  0.4 \\
$\Delta$[M/H] $=-0.15$    &  0.3 &  0.0 & $-0.3$ \\
\hline
$\sigma_{\mathrm{parameters}}$       & $\pm$1.0 &  $\pm$0.5 &  $\pm$0.7 \\
\hline
\end{tabular}
\label{mguncertainty}
\end{table}

As our final uncertainty of the isotopic fractions we add in quadrature the effects from atmospheric uncertainties to the standard error of the mean. As can be seen in Table~\ref{mguncertainty}, the isotopic fractions are not very sensitive to changes in the atmospheric parameters, highlighting one of the advantages of using Mg isotopic abundances as opposed to elemental abundances, which often show a higher parameter sensitivity. 

\section{Results}

In the 1D analysis, we were able to use all four MgH features to estimate the isotopic fractions for most stars in our sample. In a few cases we had to discard one or two of the features, due to emission being present, making them useless for the isotope derivation. This was the case for stars 20885 (two features), 28965 (one feature) and 29861 (one feature). 

\begin{figure}
\centering
\includegraphics[width=\columnwidth, trim = 7.5cm 7.5cm 0cm 0cm]{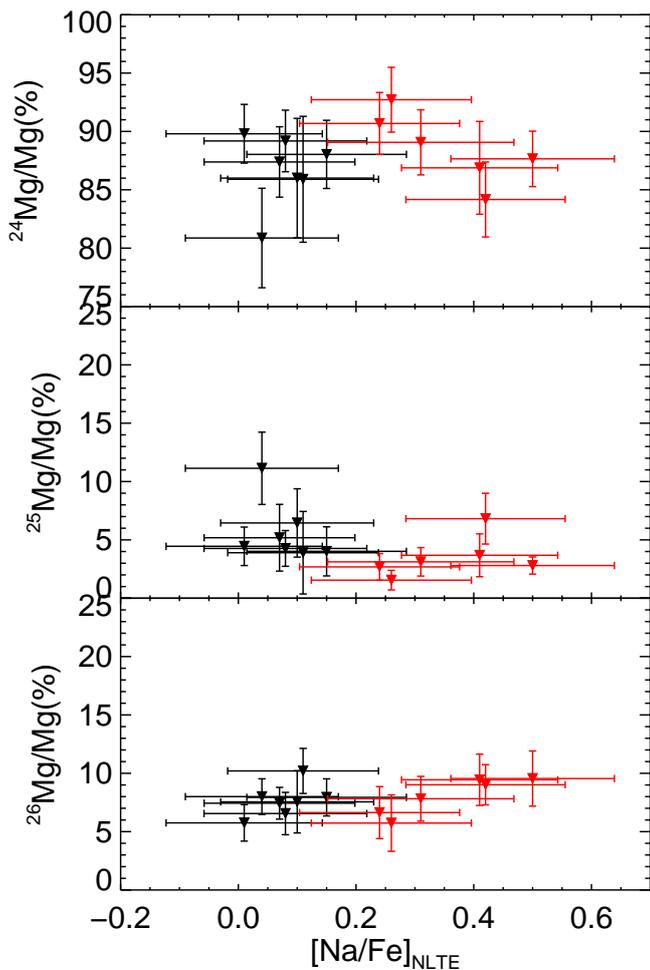}
\caption{Percentage of \mgi\ vs. [Na/Fe](top),  \mgii\ vs. [Na/Fe](middle) and \mgiii\ vs. [Na/Fe](bottom). Black triangles indicate the pristine populations and red triangles the polluted population.}%
\label{ownmg24}%
\end{figure}

In Fig.~\ref{ownmg24} we present the results from our 1D analysis, plotting the percentage of the three isotopes of the total Mg abundance vs. [Na/Fe]. In the absence of a clear Mg-Al anti-correlation, we initially plot the isotopic fractions against [Na/Fe] rather than [Al/Fe], as this allows a clear separation between the stellar populations in 47Tuc. We use the same separation criterion as in Paper 1 to distinguish between pristine (black triangles) and polluted (red triangles) stars, namely $2\sigma$ above the mean value of [Na/Fe] for field stars at the same [Fe/H]. As this criterion is somewhat arbitrarily chosen, we checked our results also using a $1\sigma$ or a $3\sigma$ selection criteria, but this did not change any conclusions presented in Sect.~\ref{discuss}. The results from each of the fitted features and the weighted means, are presented in Table~\ref{results_1d_iso}. 

In Table~\ref{summary1d} we include the [Fe/H] and the light element abundance ratios from Paper 1, together with the mean isotopic fractions, for convenience. The Na and Al abundances have been corrected for NLTE effects, whereas the Fe abundances are computed in LTE, as it was found that NLTE effects on iron were negligible, as discussed in Paper 1. In addition, we include the [C/Fe] estimates from the fitting of the C$_2$ features around 5635\,\AA, but stress that these values represents upper limits only. We find a mean [C/Fe] of $-0.03\pm0.05$ dex and $-0.35\pm0.04$ dex for the pristine and polluted population respectively. This is in excellent agreement with \citet{1997AJ....114.1051B} who fit their sample of CN-weak (pristine) and CN-strong (polluted) red giants with [C/Fe]\,$=0.00$\,dex and [C/Fe]\,$=-0.30$\,dex respectively. In particular, \citet{1997AJ....114.1051B} fails to find any significant evolution of the carbon abundance along the RGB, and rule out the strong C-depletion observed in more metal-poor clusters (see also \citealt{2001AJ....122.2561B}). Our results are also in good agreement with the study of \citet{carretta} who analysed a number of sub-giants and found [C/Fe]$\,=-0.13\pm0.02$ and [C/Fe]$\,=-0.34\pm0.04$ for the pristine and polluted populations respectively. This also reinforces the interpretation that the carbon abundance does not change appreciably during the ascent of the RGB in 47 Tucanae.

In Tables~\ref{results_1d_iso} and~\ref{summary1d}, bold face names indicate stars belonging to the polluted population. All abundances are quoted relative to the \citet{asplund} Solar abundances.

\begin{table*}%
\centering
\caption{Mg isotopic fractions for individual features, given as percentages $^{24}$Mg:$^{25}$Mg:$^{26}$Mg. Also shown are the weighted means and associated uncertainties. Boldface IDs indicate polluted stars.}
\begin{tabular}{rccccccc}
\hline
\input{mg-isotopes3.tex}
\end{tabular}
\label{results_1d_iso}
\end{table*}

\begin{table*}%
\centering
\caption{Summary of the mean Mg isotopic fractions, as well as stellar parameters and light element abundances. Boldface IDs indicate polluted stars. [C/Fe] should be taken as an upper limit only. The [Al/Fe] and [Na/Fe] abundances have been corrected for NLTE.}
\begin{tabular}{rcccccccccc}
\hline
\input{mg-isotopes-avg3.tex}
\end{tabular}
\label{summary1d}
\end{table*}

We note that one star (10237) appears to show an anomalously high fraction of \mgii, compared to the rest of the sample. This will be discussed in more detail in Sect.~\ref{outlier}

\subsection{Results from 3D}
Due to the lack of suitable 3D atmospheric models we were unable to investigate 3D effects for our full sample of stars. It is, however, still possible to investigate the importance of the improved spectral synthesis for the stars 4794, 13396 and 29861. These stars have parameters that are relatively close to the parameters of our 3D models. We refer to Table \ref{summary1d} for the stellar parameters. Because of an emission spike in the 5135\,\AA\ feature for star 29861, we were only able to derive isotopic ratios from the 5138\,\AA\ feature. 

Due to the issues with carbon blends mentioned earlier, we cannot directly compare the overall values of the Mg isotopic fractions from 1D and 3D, since we only synthesise the 5135\,\AA\ and 5138\,\AA\ features in 3D. But even just comparing the 1D and 3D results for these two features is informative to assess the potential impact of 3D.

\begin{figure*}%
\includegraphics[width=\textwidth, trim = 2cm 3cm 2cm 4cm]{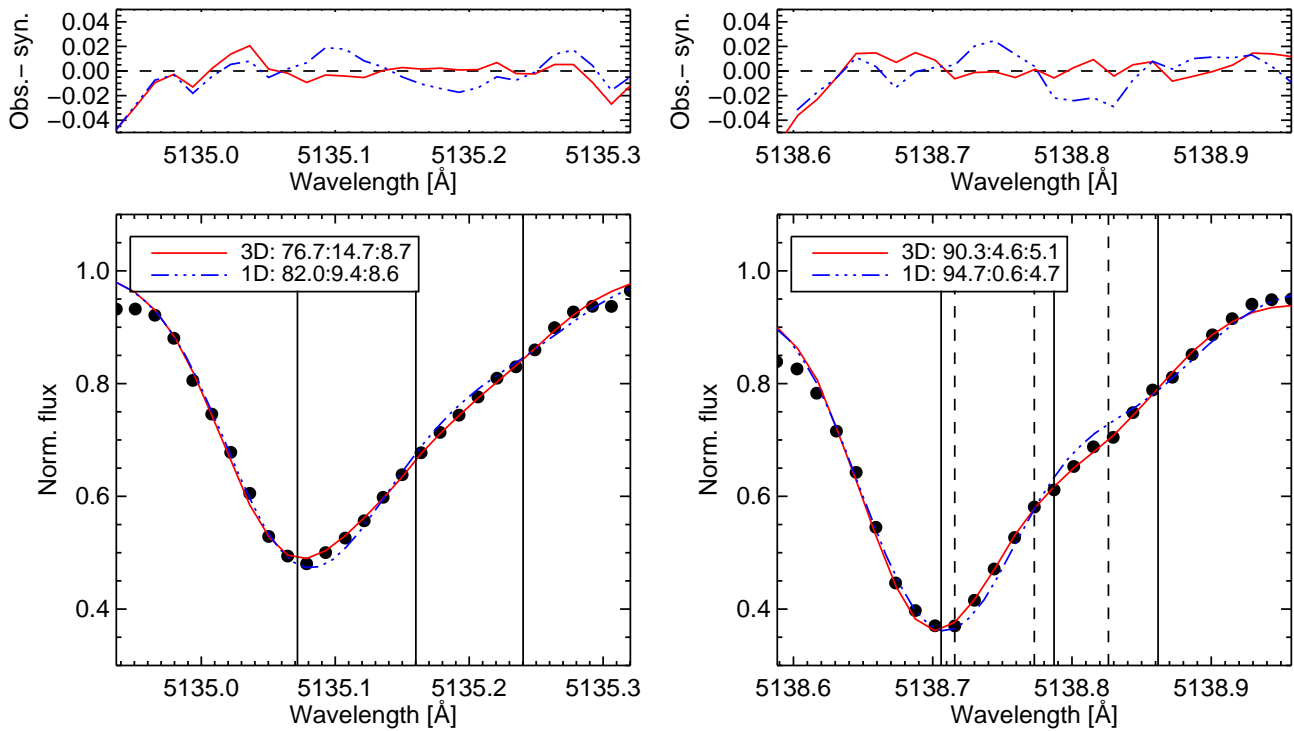}%
\centering
\caption{Best-fitting synthesis from 3D (red, solid line) and 1D (blue, triple-dot dash line) for the star 4794. Also shown are the residuals of the two fits, as well as the central positions of the MgH features.}%
\label{3dfit}%
\end{figure*}

In Fig.~\ref{3dfit} we plot the best-fitting syntheses for the two features, for star 4794. It is evident that the 3D syntheses are reproducing the observations better than the 1D syntheses. In particular, the fitting of the asymmetric wings used to derive the isotopic ratios is improved. Typical precision of the fitted percentage of each isotope are 0.7\%, 0.4\% and 0.3\% for \mgi, \mgii\ and \mgiii\ respectively. However, inspecting the distribution of isotopes that gives the best fit, in 1D and 3D respectively, it is clear that they provide rather different results, as seen in Table~\ref{3dresults}. In particular disagreement is seen for the 5135\,\AA\ feature. Here, the 3D results give a higher fraction of \mgii. This is also the case for the 5138\,\AA\ feature, although the increase, relative to the 1D results, is smaller. The \mgiii\ fraction, on the other hand, remains essentially unchanged. Comparing the mean value of the results from these two features in 1D and 3D, respectively, the 3D results suggest a higher abundance of \mgii, than what was previously reported. The feature-to-feature agreement, on the other hand, stays about the same both in the 3D and 1D synthesis. As can also be seen from Table~\ref{3dresults}, the same tendency of a significant increase in \mgii\ is observed also in the two other stars. 

\begin{table*}%
\centering
\caption{Mg isotopic ratios \mgi:\mgii\mgiii\ derived from 1D and 3D syntheses respectively.}
\begin{tabular}{lrrr}
\hline
ID & 4794 & 13396 & 29861 \\
& \mgi:\mgii:\mgiii & \mgi:\mgii:\mgiii & \mgi:\mgii:\mgiii \\
\hline\hline
1D$_{5135\angstrom}$  & 82.0:9.4:8.6  & 81.2:12.3:6.5	& Discarded \\
3D$_{5135\angstrom}$  & 76.7:14.7:8.7 & 73.5:19.0:7.5 & Discarded \\
1D$_{5138\angstrom}$  & 94.7:0.6:4.7  & 94.3:0.0:5.7  & 90.7:5.0:4.3 \\
3D$_{5138\angstrom}$  & 90.3:4.6:5.1  & 84.6:11.1:4.3 & 86.6:9.2:4.2 \\
Mean$_\mathrm{1D}$    & 88.3:5.0:6.7  & 87.7:6.2:6.1  & 90.7:5.0:4.3 \\
Mean$_\mathrm{3D}$    & 83.5:9.6:6.9  & 79.0:15.1:5.9 & 86.6:9.2:4.2 \\
$\sigma_\mathrm{1D}$  & 6.4:4.4:2.0   & 6.6:6.2:0.4   &  \\
$\sigma_\mathrm{3D}$  & 6.9:5.0:1.8   & 5.5:4.0:1.6   &  \\
\hline
\end{tabular}
\label{3dresults}
\end{table*}

\section{Discussion}	
\label{discuss}

\begin{table}%
\centering
\caption{Median value, interquartile ranges and $\sigma_\mathrm{IQR}$ for all three isotopes. Shown for the total sample and split in populations.}
\begin{tabular}{lccc}
\hline
 & median & IQR & $\sigma_\mathrm{IQR}$ \\
\hline\hline
\mgi$_\mathrm{prist.}$   & 87.7 & 3.2 & 2.5 \\
\mgi$_\mathrm{poll.}$    & 88.4 & 3.8 & 2.1 \\
\mgi$_\mathrm{all}$      & 87.8 & 3.2 & 2.4 \\
\mgii$_\mathrm{prist.}$  & 4.4 & 1.2 & 1.7 \\
\mgii$_\mathrm{poll.}$   & 3.0 & 1.0 & 0.9 \\
\mgii$_\mathrm{all}$     & 4.0 & 2.1 & 1.4 \\
\mgiii$_\mathrm{prist.}$ & 7.5 & 1.4 & 1.3 \\
\mgiii$_\mathrm{poll.}$  & 8.4 & 2.8 & 1.5 \\
\mgiii$_\mathrm{all}$    & 7.7 & 2.6 & 1.4 \\
\hline
\end{tabular}
\label{mgspread}
\end{table}

The presence or absence of correlations between the Mg isotopes and the abundance ratios of the light elements can provide insight into the underlying mechanism responsible for the abundance variations. 

\begin{figure}%
\centering
\includegraphics[width=\columnwidth, trim = 7.5cm 7.5cm 0cm 0cm]{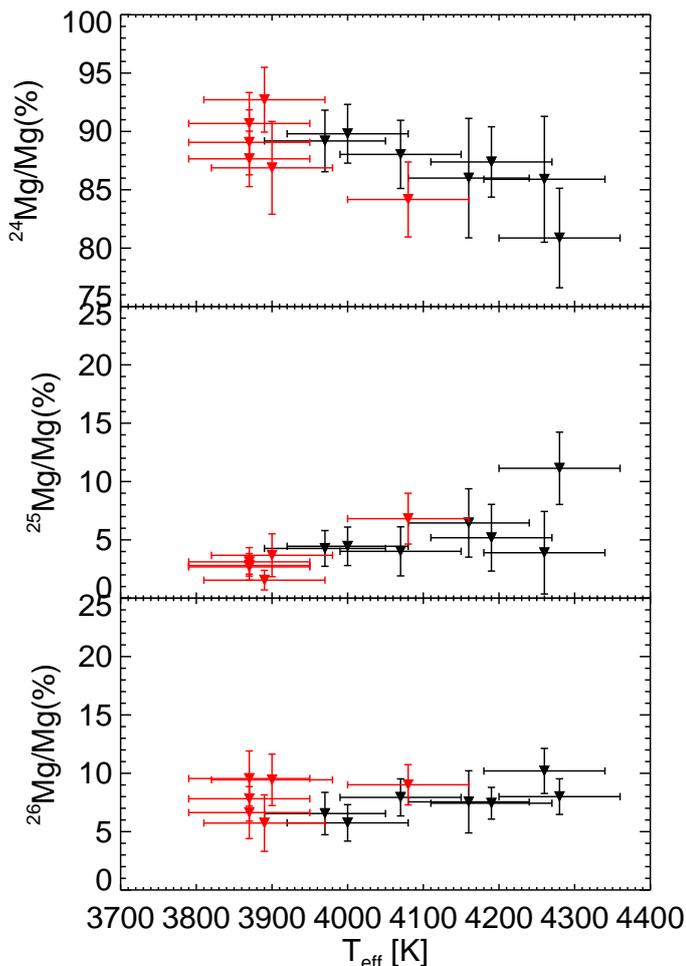}%
\caption{Percentage abundances of the Mg isotopes vs. \teff.}%
\label{teffcorr}%
\end{figure}

As illustrated in Fig.~\ref{teffcorr}, our \mgi\ and \mgii\ isotopic fractions appear to show some correlation with \teff. The trend is fairly weak for both affected isotopes, in particular if the outlier that appears strongly depleted in \mgi\ and enhanced in \mgii\ is discarded (star 10237). By doing so, only the correlation between \mgii\ and \teff\ has a slope that is different from zero by more than $2\sigma$. There may be some justification for not considering this star, as will be discussed later. Our \mgiii\ measurements, on the other hand, show no trend with \teff. Since the formation properties are identical for all three isotopic components of MgH, we attribute this weak parameter correlation to blends from other lines, which will influence the isotopic components differently.

As discussed in Sect.~\ref{lineselection}, the blending C$_2$ lines in the 5134\,\AA\ and 5140\,\AA\ features are temperature sensitive. As such, the inclusion of these blending lines can introduce a temperature dependent systematic effect on the derived isotopic ratios . Indeed, if we consider only the Mg isotopic fractions from the 5135\,\AA\  and 5138\,\AA\ features, our results do not show any significant correlation with \teff\ for any of the three isotopes. As such, we do not take the apparent correlation as an indication of a parameter problem with our stars, but rather a problem with the blending lines. This issue was also addressed in Paper 1, where we discussed correlations of some abundances with the stellar parameters. The initial abundances of [Ba/Fe] and [Al/Fe] showed a trend with temperature, but this was found to be fully explained by NLTE effects on Ba and Al. Nevertheless, the fact that the bulk of the polluted stars happen to be the coolest giants in our sample, makes it difficult to disentangle possible systematics due to a parameter problem, from genuine differences in abundances between the stellar populations, but we note that we do see a range of isotopic ratios for a given \teff. In addition, the temperature differences between the stars belonging to the pristine and polluted populations are fully consistent with their positions in the colour-magnitude diagram (Fig.~1 in Paper 1). As such, we do not interpret the weak parameter correlation to be due to a problem with the \teff\ scale, but rather due to inaccurate line data for the blending lines.

In Fig.~\ref{m71comp} we present the measured isotopic fractions vs. [Al/Fe]. This is a natural choice, since it is expected that the heavy isotopes are created together with Al. It is clear from the figure that we do not observe a broad spread in the Mg isotopes in either of the two populations of stars. To inspect the spread in our data we compute interquartile ranges (IQR) and median values, both for our sample as a whole, as well as separately for the two populations (Table~\ref{mgspread}). We choose the IQR, rather than the usual standard deviation as a measure of spread, since the former is better suited for small samples, whereas the standard deviation is much more prone to being skewed by outliers in the data. In the case of a normal distribution of measurements, the IQR extends to $\pm0.7\sigma$ from the mean, where $\sigma$ is the normal Gaussian standard deviation. Since we assume Gaussian uncertanties, we scale the reported $1\sigma$ uncertainties by this value, and call it $\sigma_\mathrm{IQR}$. If we measure an IQR significantly higher than this, we can claim an intrinsic spread in our data, assuming that all systematics are accounted for. These values are reported in Table~\ref{mgspread}. 

From this, it is clear that the sample as a whole, does not show any significant spread for any of the Mg isotopes. Indeed, the IQR is comparable to $\sigma_\mathrm{IQR}$ in all cases. This also holds true when inspecting the pristine and polluted populations separately, where no significant spread is observed in either population. Furthermore, the median values for all isotopes are in agreement within the typical uncertainties on the measurements. As such, we do not see any evidence for a different behavior of the Mg isotopes in the two populations of stars. The only weak indication that may be present is that the polluted population shows a spread in \mgiii\ that is twice that of the pristine population, but the spread is not statistically significant and is likely just reflecting our small sample size.

Inspecting the plot of our Mg isotopic fractions against [Al/Fe] in Fig.~\ref{m71comp}, there seems to be a trend of \mgi\ and \mgiii\ with the Al abundance, particularly evident for the polluted population. To investigate the strength of any potential trend, we performed linear fits to the observed values, taking into account uncertainties on both [Al/Fe], and on the individual Mg isotopic fractions. We used the \texttt{MPfitexy} routine, described in Sect. 4 of \citet{williams}, which builds on the \texttt{MPfit} package of \citet{markwardt}. Linear fitting was done both using the sample as a whole, as well as when split into the pristine and polluted populations sub-samples. We consider the fitted trend significant if the resulting slope is different from zero by more than $2\sigma_\mathrm{slope}$, with $\sigma_\mathrm{slope}$ being the uncertainty on the best-fitting value of the line slope. For none of the fitted populations were we able to detect a significant trend of the Mg isotopes with [Al/Fe]. We performed the same test using only the average of the results from the 5135\,\AA\ and 5138\,\AA\ features, to check whether the inclusion of the MgH features with strong C$_2$ blends had any influence. Also in this case we did not detect any significant correlation between any of the isotopes and [Al/Fe].

\subsection{Star 10237}
\label{outlier}
An outlier in our sample is the star 10237. It appears depleted in \mgi, $2\sigma$ below the sample mean. It is also found to be enhanced in \mgii, being $\sim3\sigma$ above the mean value for the full sample. It is also the star with the highest derived \teff\ in our sample. This behaviour is rather striking and there can be a number of explanations. In particular the \mgii\ abundance is surprising, as this has not been seen for any other star in our sample, and even compared to the remaining GCs with Mg isotope measurements, it is amongst the highest values observed. 

One possible explanation could be that this star is in a binary system, where mass-transfer has occurred in the past, and thus it got enhanced in \mgii\ from material accreted from a companion AGB star. However, in this case, an enhancement in \mgiii\ would also be expected, which we do not observe. With only one epoch of observation, we cannot tell whether this star is member of a binary system. A potential companion white dwarf would not make itself known by double lined spectra, and in any case we do not see any evidence for this. In this case, we would also expect to see other signatures of AGB nucleosynthesis, which we do not. The expected signature, does naturally depend on the yield of the adopted AGB model, and the assumptions about dilution. For instance the models of \citet{ventura} do not predict a very strong oxygen depletion, although Na can get enhanced by almost an order of magnitude.  

Since only massive AGB stars reach temperatures high enough to activate the Mg-Al burning chain, we rule out self-pollution as a potential explanation. At the age of 47 Tucanae, stars with $M\gtrsim4$\msun\ will already have evolved well beyond the AGB stage. 

This star shows the lowest magnesium abundance of all stars in our sample ([Mg/Fe] = 0.32 dex), which makes the MgH features weaker, compared to any other star in our sample. The high abundance of \mgii\ seems robust, though, as all MgH features indicate this, albeit with some scatter between the different features, as is also seen for the remaining stars. One could speculate that there might be undetected blends that only become important at these higher temperatures, but considering the excellent agreement of the isotopic fractions between the MgH features, this does not seem to be a reasonable explanation. Blends that are not accounted for would result in an over-estimation of the Mg abundance. Thus, if this was the case, the measured abundance of Mg would become even lower.

Another option could be that this star is in fact not a member of the cluster. Whereas both the position in the CMD and the radial velocity reported in Paper 1 is consistent with membership, we may see indications in the abundances that this is an interloper. Considering that this star has the highest [O/Fe] of all stars in our sample, and that SN II may have contributed with some heavy Mg isotopes in the field at this metallicity, this could indicate that this star has been polluted by more supernova material than the remaining sample. This star also has one of the highest [Eu/Fe] values of the stars in our sample. Indeed comparing our results to the investigation of Mg isotopes in field stars by \citet{yong_field}, the average isotopic fractions at this metallicity is 80:10:10, which compares well to what we observe for this star (80.9:11.8:7.3). The isotopic fraction of heavy Mg in the field stars is also significantly higher than what is seen for the rest of our cluster stars, at this metallicity. Thus, we cannot rule out that this star may be a non-member.

\subsection{Comparison with earlier works}
Several other studies have also investigated Mg isotopes in globular clusters, and used these results to constrain the nature of the intra-cluster polluters (NGC6752, \citealt{yong6752}; M13, \citealt{shetrone,yongm71}; M71, \citealt{melendezm71}, hereafter MC09 and $\omega$Cen, \citealt{dacosta}).

The most obvious candidate for a direct comparison is M71, as it has a nearly identical metallicity to 47Tuc ([Fe/H] $=-0.72$ dex, \citealt{melendezm71}, hereafter MC09). These authors separated their sample into CN-strong and CN-weak stars and inspected the Mg isotopes for these populations. They found that the CN-strong stars were depleted in \mgi\ and enhanced in \mgiii, relative to the CN-weak stars, but still found that the overall variation was rather small, albeit slightly higher than what we see in our sample. 

\begin{figure}%
\centering
\includegraphics[width=\columnwidth, trim = 7cm 7cm 0cm 0cm]{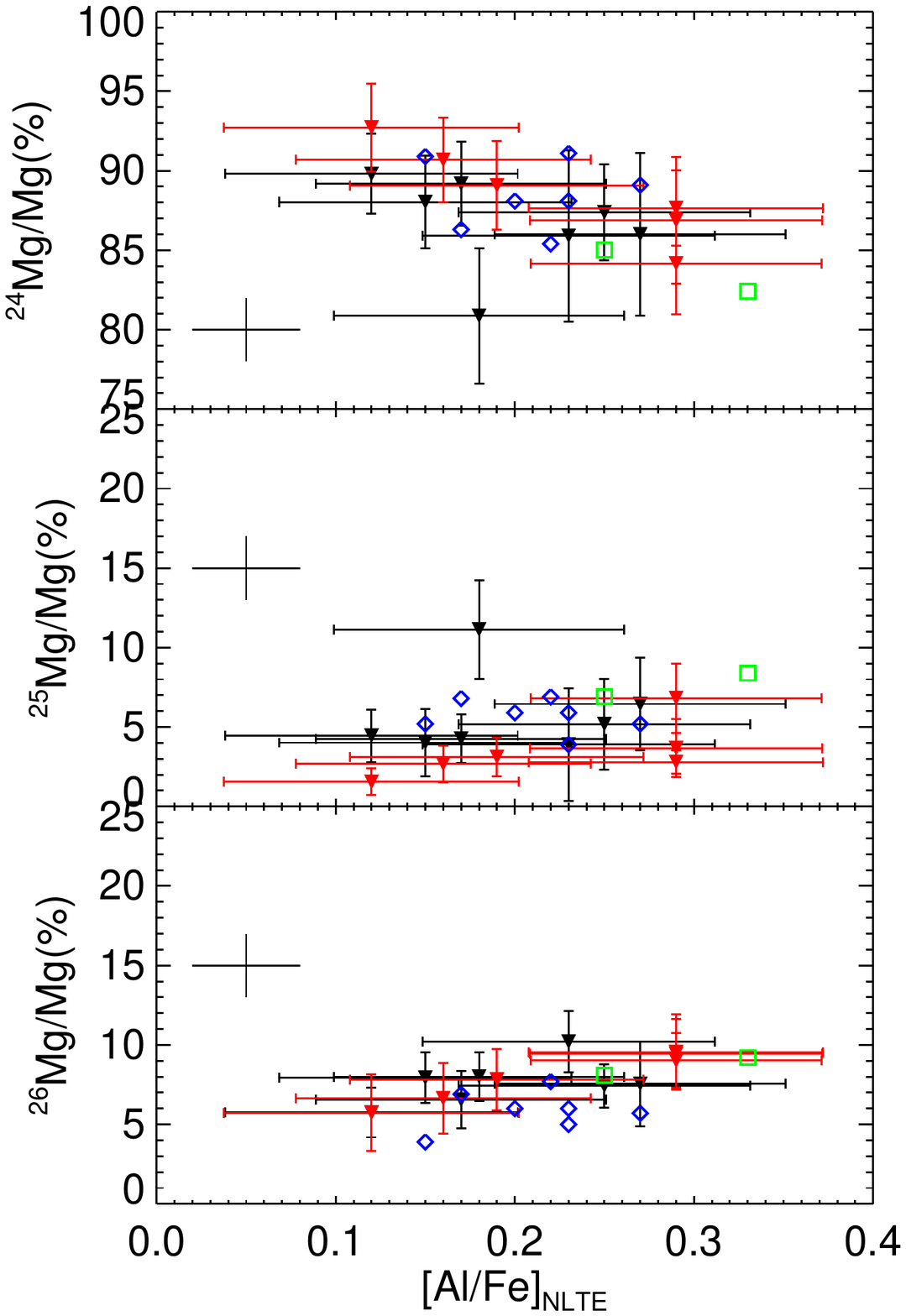}%
\caption{Percentage of \mgi\ vs. [Al/Fe] (top), \mgii\ vs. [Al/Fe] (middle) and \mgiii\ vs. [Al/Fe] (bottom). Black triangles indicate the pristine populations and red triangles the polluted population. Also shown are the results from \citet{melendezm71} for M71. In the M71 sample, the polluted population is indicated with green, open diamonds. Blue open diamonds show the pristine population. Typical uncertainties of the isotope measurements for M71 is also shown.}%
\label{m71comp}%
\end{figure}

To allow for a better comparison between the studies, we adopt the same population separation criterion, based on the [Na/Fe] value for M71. Before making the separation, we shift the results of MC09 to the abundance scale of \citet{asplund}. We use the same [Na/Fe] value to split the M71 population as we used in Paper 1. In Fig.~\ref{m71comp} we plot our measurements, together with the ones of M71, when using the same population discriminant, as a function of [Al/Fe]. Clearly, the behaviour of the isotopes is very similar to what we see in 47Tuc. A small variation is seen in \mgi\ and \mgiii, and an approximately constant value of \mgii. The MC09 stars appear to be slightly more enhanced in \mgii, compared to ours. The small differences found can likely be attributed to differences in the choice of MgH features, where MC09 used the 5134.3\,\AA\ and not the 5134.6\,\AA\ feature that we are considering. They did also not use the 5135.3\,\AA\ feature, which we chose to include. Differences in the selection of blending atomic lines will also play a role. In addition, we do not expect the chemical evolution of the two clusters to have been exactly the same, since 47Tuc is significantly more massive than M71. It is worth noting that when splitting the MC09 sample based on their [Na/Fe] values, only two stars are found to belong to the polluted population of stars. These two, however, do show the strongest \mgi\ depletion and \mgiii\ enhancement in the MC09 sample, suggesting that these stars are made up of a greater fraction of processed material, compared to the remaining stars in the sample. A third star in the MC09 sample falls very close to our cut between the two populations, and could in principle belong also to the polluted population. This star shows the highest fraction of \mgii\ and \mgiii\ of the pristine stars, so it would not change the conclusion about the polluted population in the MC09 sample. 

\begin{figure}%
\centering
\includegraphics[width=\columnwidth, trim = 2cm 3cm 2cm 3cm]{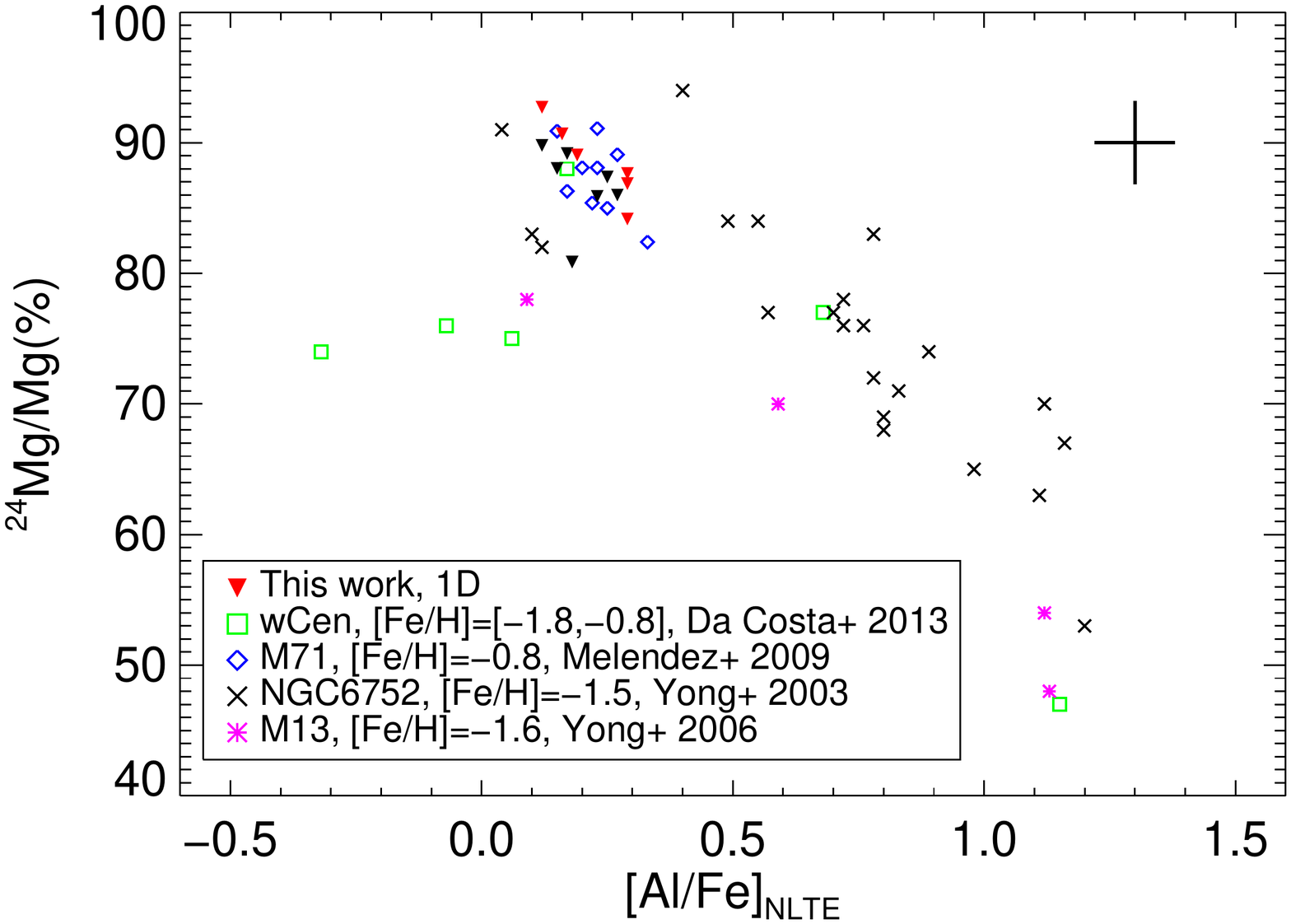}%
\caption{Linear fraction of \mgi\ vs. [Al/Fe]. Our results are shown as black and red triangles. For comparison the results from \citealt{dacosta} (green squares), \citealt{melendezm71} (blue diamonds), \citealt{yong6752} (black crosses) and \citealt{yongm71} (pink asterisks) are shown.}%
\label{resmg24}%
\end{figure}

\begin{figure}%
\centering
\includegraphics[width=\columnwidth, trim = 2cm 3cm 2cm 3cm]{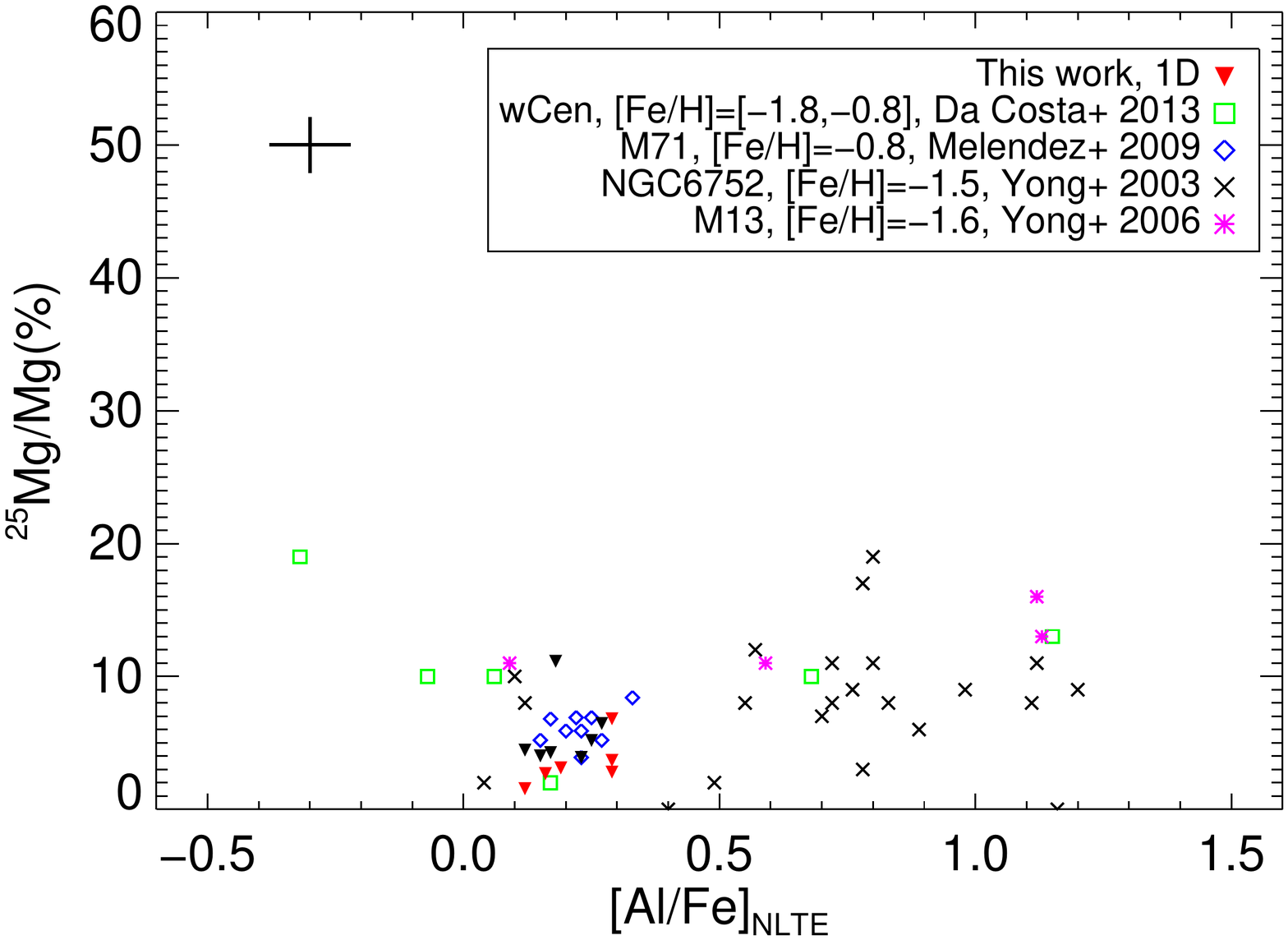}%
\caption{Same as in Fig.~\ref{resmg24}, but for \mgii.}%
\label{resmg25}%
\end{figure}

\begin{figure}%
\centering
\includegraphics[width=\columnwidth, trim = 2cm 3cm 2cm 3cm]{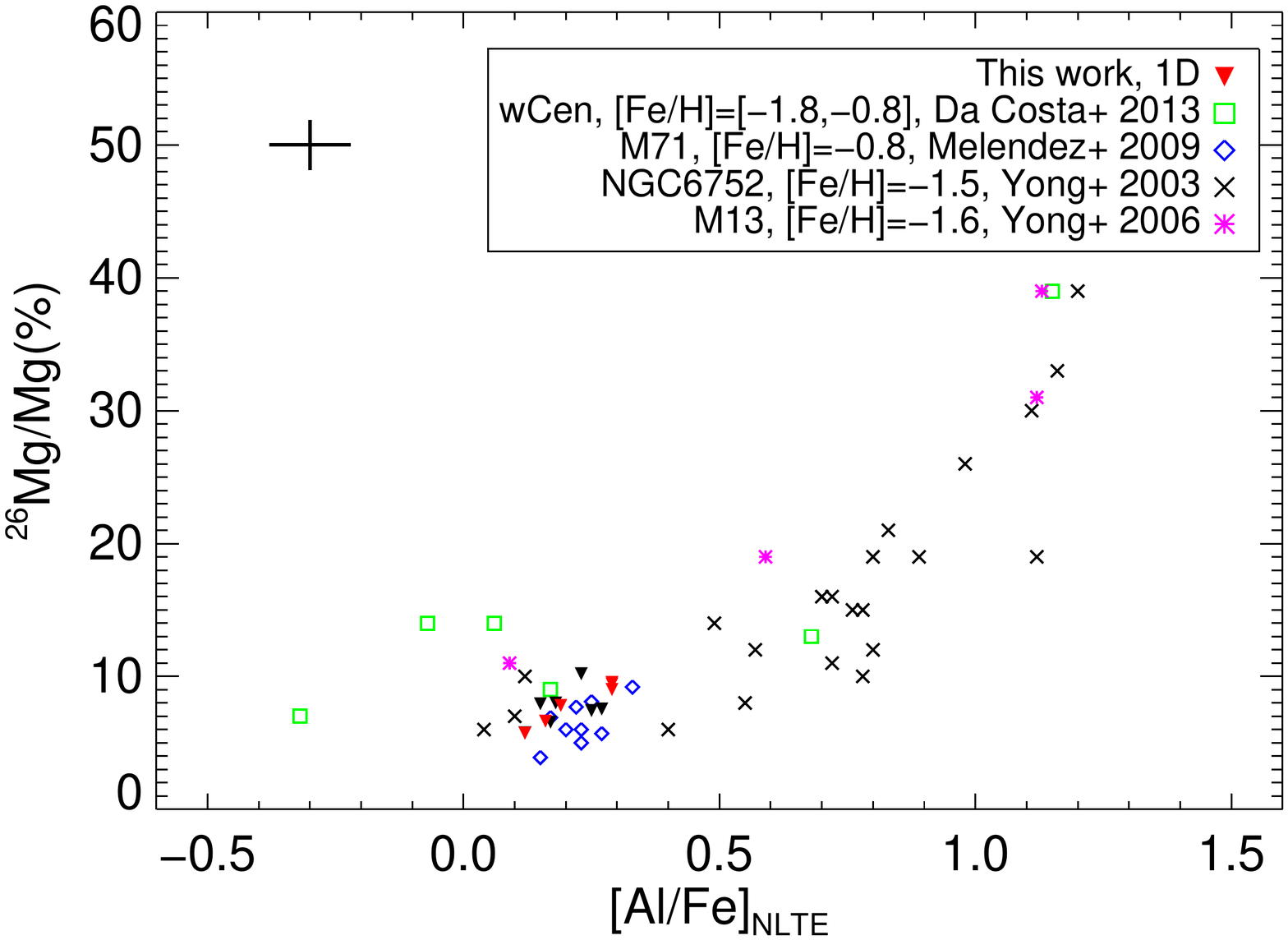}%
\caption{Same as in Fig~\ref{resmg24}, but for \mgiii.}%
\label{resmg26}%
\end{figure}

Comparing now to the remaining studies of Mg isotopes in GCs, which are shown in Figs.~\ref{resmg24}, \ref{resmg25} and \ref{resmg26}, our results fall nicely on the trend shown by these works. It is evident that the amount of \mgi\ only starts to change appreciably, once the cluster has seen a significant activation of the Mg-Al burning chain. Until [Al/Fe] reaches approximately 0.5 dex, there is no significant variation in the isotopes. Only for more aluminum-enhanced stars do we begin to see a strong depletion of \mgi\ and a simultaneous increase in the \mgiii\ abundance. The only exception here, being the results from $\omega$ Centauri, where all stars appear depleted in \mgi. However, this cluster also shows a spread in metallicity, and the stars analysed do indeed have different [Fe/H] values, so they have likely had a more complicated formation history, compared to the rest of the GCs. It is also worth noting that even for the clusters with the widest ranges in [Al/Fe] (NGC6752 and M13), the abundance of \mgii\ stays approximately constant. We note that since we employ one new feature, compared to the remaining studies, there may be a small offset between our results and the comparison studies. However, considering the overall good agreement with other clusters, for the same [Al/Fe] values, we do not expect this to have a strong impact.

Assuming that AGB stars are the main source of heavy Mg isotopes, the lack of variation in the isotopes in 47Tuc, may be linked to its higher metallicity, which lowers the temperature at the bottom of the convective envelope, where HBB occurs in AGB stars. This, in turn, results in less efficient nucleosynthesis compared to the low metallicity cases. This may also be linked to the smaller Al-variations, often seen for high-metallicity clusters. However, we do not sample the full range of Al abundances in 47Tuc. \citet{cordero} reports a total range of $\sim0.7$\,dex and \citet{carretta2013} find a $\sim0.5$\,dex range in [Al/Fe], which should be compared to our total range of 0.17 dex. Thus, we do not expect our sample of stars to cover the most extreme chemistry stars in 47Tuc. Even including NLTE effects will not change this picture, as \citet{cordero} find stars with near-identical parameters having [Al/Fe] abundances that differ by more than 0.4 dex. Similar differences are seen by \citet{carretta2013} for stars with identical parameters. These stars will have identical NLTE corrections, so as such, the variation in Mg isotopes in 47Tuc may be larger than what reported here, as the most Al-enhanced stars should coincide with the extremes of the heavy Mg isotope fractions. 

Relatively large spreads in [Al/Fe] are also seen in other high-metallicity clusters like Terzan 5 ($\Delta$[Al/Fe]$\,=0.47$\,dex for both the low and high metallicity population, $\Delta$[Al/Fe]$\,=0.87$\,dex for the total sample, \citealt{2011ApJ...726L..20O}), M71 ($\Delta$[Al/Fe]$\,=0.50$\,dex, \citealt{carretta2009}) and NGC 6553, ($\Delta$[Al/Fe]$\,=0.37$\,dex \citealt{2014AJ....148...67J}). Again, stars with identical parameters show differences of up to 0.47\,dex (Terzan 5), 0.42\,dex (M71), and 0.35\,dex (NGC 6553), so the differences are not due to NLTE effects on Al, and one might expect these clusters to also show variation in the heavy Mg isotopes, as already shown for M71 by MC09.

\subsection{Pollution scenarios}

\begin{figure}%
\centering
\includegraphics[trim = 2cm 4cm 5cm 4cm, width=\columnwidth]{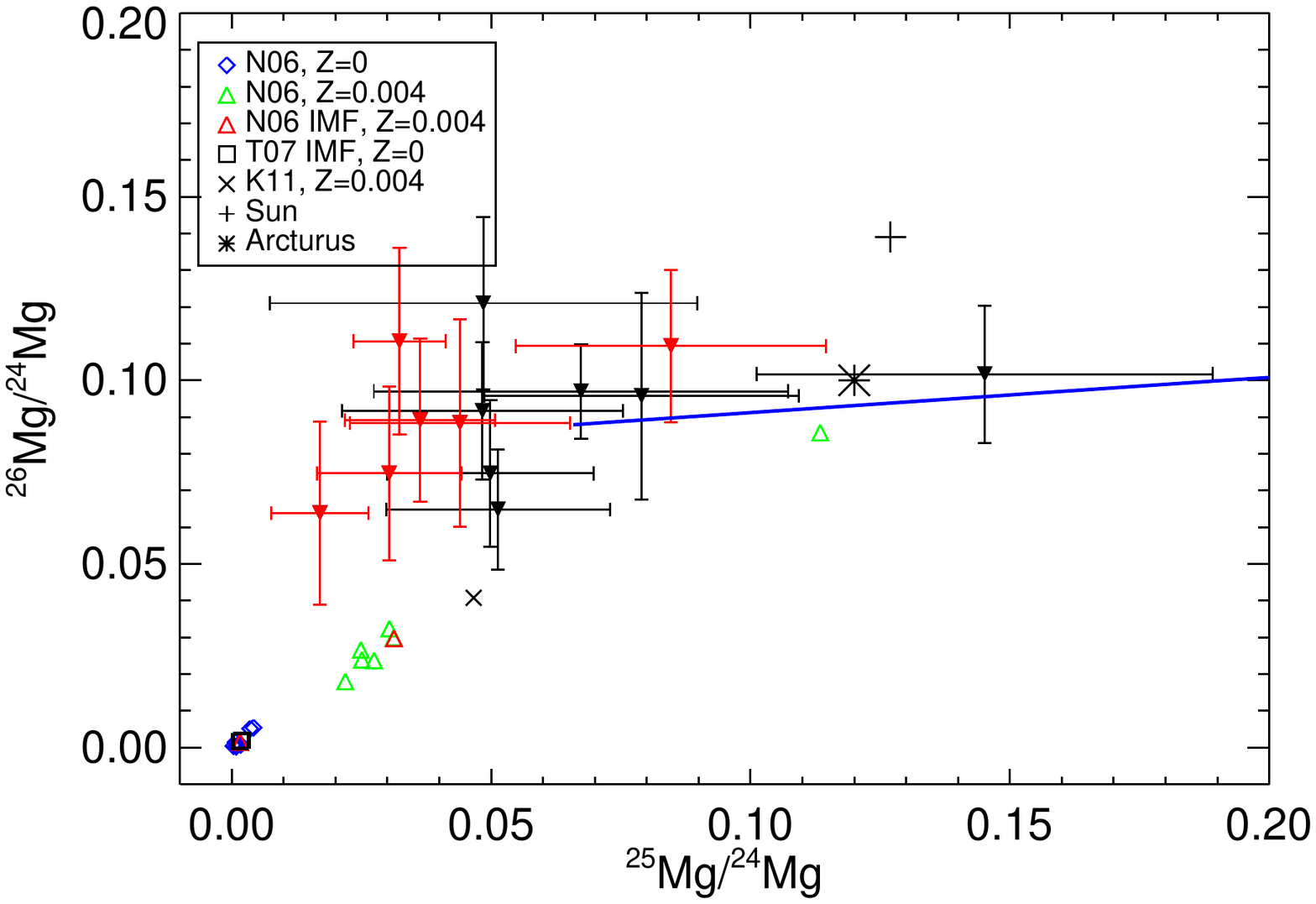}%
\caption{\mgiii/\mgi\ vs. \mgii/\mgi\ measured in our sample. Shown are also isotopic ratios from the supernova yields of \citet{nomoto} (Z=0, blue diamonds; Z=0.004, green triangles; Z=0.004 IMF weighted value, red triangle), \citet{tominaga} (Z=0, black squares) and from the chemical evolution model of \citet{kobayashi} (black X). Shown is also the dilution curve for the isotopes using the model form \citet{ventura}. The solar position is shown with a black cross, and our values for Arcturus with a black asterisk.}%
\label{ratios}%
\end{figure}

In Fig.~\ref{ratios} we present our measured isotopic ratios. It is evident that both population exhibit both low \mgii/\mgi\ ratios and low \mgiii/\mgi\ ratios, and that there is no significant difference between the two populations. 

In the figure, we also present the predictions from a number of supernova models. From \citet{nomoto}, we present yields from a range of progenitor masses at both $\mathrm{Z}=0$ and $\mathrm{Z}=0.004$, corresponding to the metallicity of 47Tuc. We also indicate their average yields, weighted by the initial mass function (IMF). In addition, we show results from \citet{tominaga}, but here only given as the IMF-weighted values. Finally, we show the isotopic ratios predicted for the Milky Way halo from the chemical evolution study of \citet{kobayashi}. As a reference, also the isotopic ratios of the Sun and Arcturus, are indicated with a black cross and a black asterisk respectively.

The results presented here, together with the results of Paper 1, allow us to investigate potential pollution scenarios in more detail. Whereas our sample of stars does not cover the entire range in [Al/Fe] of the cluster (\citet{cordero} and \citet{carretta2013} find stars with identical parameters that differ by up to 0.43 dex in [Al/Fe], more than twice the range we find), we can still gain insight into its chemical evolution. 

Heavy isotopes of magnesium can form from a number of different nucleosynthesis processes. It is indeed possible for supernovae of massive stars to form \mgii\ and \mgiii\ through $\alpha$-captures on Ne, in the $^{22}\mathrm{Ne}(\alpha,n)^{25}\mathrm{Mg}$ and $^{22}\mathrm{Ne}(\alpha,\gamma)^{26}\mathrm{Mg}$ processes (see e.g. \citealt{kobayashi}). This formation channel requires an initial seed of Ne, and is thus found to increase with increasing metallicity of the supernova progenitors. Up to a metallicity of about [Fe/H$]=-1.0$ dex, no significant production is expected from this channel \citep{alibes,prantzos,kobayashi}. In AGB stars it is also possible to produce some heavy Mg isotopes from the above process, since some primary production of Ne is possible. However, the AGB production of heavy Mg isotopes is dominated by the Mg-Al chain during Hot Bottom Burning.

For metal-poor clusters, it is typically assumed that the proto-cluster cloud consisted of fully mixed gas, enriched by the yields of metal-free supernovae. However, it is clear from the abundance pattern of 47Tuc that the proto-cluster cloud has seen some contribution from a non-supernova source, since a clear contribution from the s-process is observed (e.g. \citealt{cordero} and Paper 1). Since s-process elements are synthesised in AGB stars, one would expect that also the Mg isotopes should be elevated to levels above that of pure supernova yields.  

Whilst the measured ratios in 47Tuc are still below the solar values of \mgii/\mgi=0.127 and \mgiii/\mgi=0.139, both the pristine and polluted populations show ratios that are higher than what is predicted from supernova yields alone. In particular, the zero-metallicity supernovae only produce trace amounts of the heavy isotopes, clearly inconsistent with our observations. This was noted already in the study of NGC6752 by \citet{yong6752}, and the same pattern is observed for other GCs with measured isotopic ratios of Mg. Even considering yields from supernovae at the metallicity of 47Tuc, there is still a marked under-production of the heavy isotopes from the SN channel alone, in particular true for \mgiii. 

It is also worth noting that even when one looks at the general Milky Way evolution, as done in the \citet{kobayashi} study, there are still not enough heavy isotopes being produced in the models. This was noted by the authors already in the original study, when comparing to measurements from field stars, where the field stars have even higher isotopic ratios than what we observe. 

In the case of 47Tuc, although one would expect some AGB contribution, as also suggested by the s-process abundances, the models of \citet{kobayashi} are still underpredicting the abundance of \mgiii\ in particular, but to a lesser extent than for field stars at the same metallicity. Our 1D results in Fig.~\ref{ratios} thus reinforce the interpretation from Paper 1 that the proto-cluster cloud has seen contributions from processes other than type II supernovae, even beyond that predicted for the general Milky Way chemical evolution. Some additional source of especially \mgiii\ appears to be needed in the models.

\subsubsection{The AGB scenario}
\citet{ventura} recently proposed a pollution scenario for 47Tuc, under the AGB scheme. They were able to reproduce the light element abundance variations in a large fraction of the observed stars from the \citet{carretta2013} sample, assuming a varying degree of dilution of the AGB ejecta with pristine material. Our results for [O/Fe] and [Na/Fe] for the polluted stars place them within the range covered by their proposed dilution curve for the abundance ratios, when the abundances are offset by $-0.2$ dex and $-0.3$ dex, respectively, as done in their work. The pristine stars, on the other hand, fall just outside the range covered by their models. We note that the AGB models of their work can only create a 0.2 dex variation in [O/Fe] without violating the constraints on the maximum He variations within the cluster, obtained from analyses of the Horizontal Branch \citep{dicriscienzo}, and the width of the main sequence \citep{milone}. Since the [O/Fe] variations span more than 0.8 dex in 47Tuc \citep{cordero}, some additional source of light element variation may be needed. A possible solution to this could also be extra mixing processes in giants, which can result in an additional depletion of oxygen in the polluted population of stars, as investigated by \citet{dantona_mixing}. This is expected only to happen in the most extreme chemistry stars, and such stars are expected to also be strongly enhanced in He. While the overall He spread in 47Tuc is small, a small population of extreme chemistry stars may exist, which is also supported by the fact that only a handful of stars with such strong oxygen depletion has been found by \eg\ \citet{cordero}.

The AGB polluters in \citet{ventura} are in a mass range where the stars experience a mild Hot Bottom Burning, which is where the Mg-Al burning chain is activated. Their models predict an enhancement of the AGB surface abundances of Al of up to +0.5 dex, whereas the overall Mg abundances are only barely touched, with a depletion of at most 0.04 dex. Considering that the pure AGB ejecta need to be diluted with gas of pristine composition, no detectable variation in the Mg abundance is expected, consistent with our results. On the other hand, one would expect a small variation in [Al/Fe] in the polluted stars, which has indeed been reported by both \citet{carretta2013} and \citet{cordero}. That we do not see any statistically significant variation in [Al/Fe] in our sample of stars is likely a consequence of our small sample size, as discussed in Paper 1.

In Fig.~\ref{ratios}, we show the dilution curve for \mgiii/\mgi\ vs. \mgii/\mgi, using the same AGB models as in \citet{ventura}. The amount of AGB material increases from left to right in the plot, with the leftmost point giving the pristine composition assumed in their models. Clearly, the bulk of our measured isotopic ratios fall well below the predicted value, even when considering a composition of purely pristine material. However, if one started from a composition with a lower abundance of heavy Mg isotopes, the curve would shift to the left, and show much better agreement with the observations. We note here that the pristine composition assumed by \citet{ventura} starts out with a higher initial fraction of the heavy Mg isotopes, which would explain this offset.

\begin{figure}%
\centering
\includegraphics[trim = 7cm 7cm 0cm 0cm, width=\columnwidth]{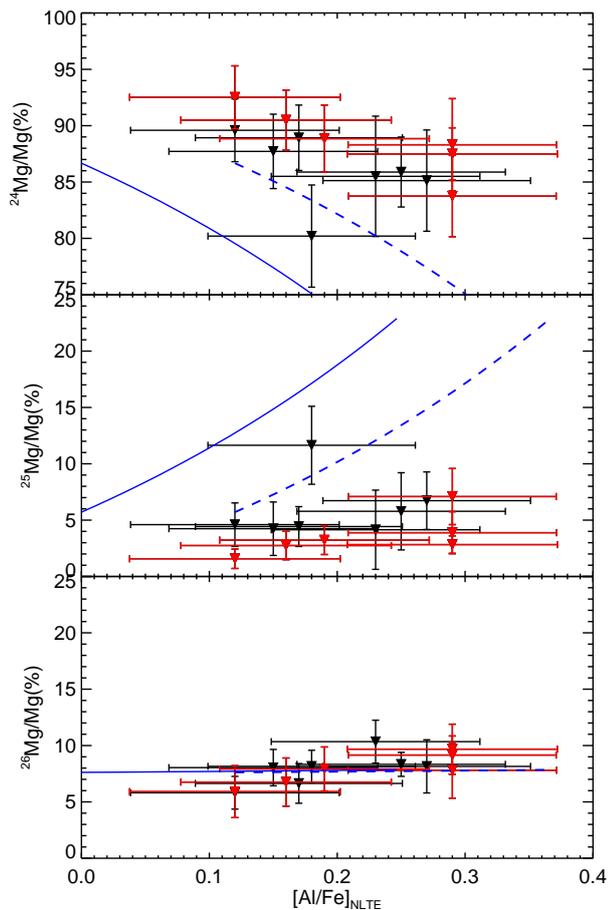}%
\caption{Isotopic fractions vs. [Al/Fe]. The solid blue lines is the predicted composition of the stars from \citet{ventura}. The dashed lines shows the dilution curve, when shifted by +0.12 dex in [Al/Fe].}%
\label{iso-al}%
\end{figure}

During the burning chain that creates the Al enhancement in AGB stars, production of primarily \mgii, but also \mgiii\ is occurring, and one would thus expect a change in the isotopic distribution between the pristine and polluted populations. Indeed, the AGB models of \citet{ventura} predict a strong decrease of \mgi\ whereas \mgii\ should increase by almost an order of magnitude. Thus, one would expect the \mgii\ abundances to correlate with [Al/Fe] for the polluted population. As discussed earlier, our analysis does not give any indication of such a correlation being present. This is further illustrated in Fig.~\ref{iso-al}, where we plot the dilution curve (solid line) for the Mg isotopic fraction vs. [Al/Fe], using the models of \citet{ventura}. The maximum allowed fraction of AGB material is 0.3, as imposed by the constraints from the He abundance variation measured from the extent of the Horizontal Branch in the cluster. Whereas our [Al/Fe] values are within the model prediction, we see no evidence for a strong increase in \mgii. Thus, our observations are somewhat in disagreement with the predictions of the AGB models of \citet{ventura}, where an increase in \mgii\ would be expected, even if the Mg-Al chain is only weakly activated. 

The dashed blue line in Fig.~\ref{iso-al} shows the position of the dilution curve, if we assume that the lowest observed value of [Al/Fe] in our pristine stars reflects the actual abundance in the pristine population, and shift the dilution curve by this amount (+0.12 dex). The agreement is now significantly better, but the models still over-produce \mgii, compared to what we observe, in particular in the [Al/Fe] enhanced region. The \mgiii\ vs. [Al/Fe] on the other hand, shows good agreement with the models which is not surprising, as the models do not predict any detectable variation in this isotope.

The lack of variation in \mgii\ is a well-known problem, also in other GCs, where the observed values of \mgii\ are found to be approximately constant across 1 dex in metallicity, and about 1.5 dex in [Al/Fe], so it is not a phenomenon reserved for high-metallicity clusters like 47 Tucanae. Furthermore a significant variation in \mgiii\ is observed for the lower metallicity clusters, which is not predicted by the AGB polluters.

\subsubsection{Other pollution mechanisms}
AGB star models have seen the most research in the GC context, but they are not the only sources capable of producing heavy Mg isotopes. Indeed, the recently proposed scenario of \citet{denissenkov} is able to simultaneously reproduce the Na-O and Mg-Al anti-correlations, and the observed variation in heavy Mg isotopes and their correlation with [Al/Fe] for the clusters with reported Mg isotope measurements. They invoke supermassive stars ($>10^4$\,\msun) to explain the abundance patterns, and these are indeed able to undergo the required nuclear burning. Their models are fully convective and processed material can easily be transported to the surface, and lost to the cluster where it is mixed with pristine material to a smaller or lesser degree. In their models, they need to shut off nuclear burning once the central He abundance has increased by $\Delta Y=0.15$ in order to not violate the maximum observed He variations in Galactic GCs. This is proposed to happen due to fragmentation of the supermassive stars. Although their model is appealing from many aspects, it is not without problems. For instance, a given extent of the Na-O anti-correlation requires a certain amount of processed material, which will also be enhanced in He. This poses a problem, since clusters with comparable Na-O variations show significantly different He variations \citep{bastian2}. Indeed, \citet{2015MNRAS.448.3314D} explore the GC M13 in some detail and from their model it is suggested that this cluster should exhibit a variation in He of $\Delta Y=0.13$, in contrast with the maximum observed variation of $\Delta Y=0.04$ found by \citet{2013MNRAS.430..459D}. 

Unfortunately, \citet{denissenkov} only presents models for a single metallicity ([Fe/H$]=-1.5$\,dex), so their predictions are not directly comparable to the case of 47 Tucanae. Nevertheless, if it is assumed that the yields of the supermassive stars scale so that the yields are proportionally the same at the metallicity of 47 Tucanae, we can still make a qualitative assessment of the feasibility of this candidate. Inspecting their predictions for the variation of the Mg isotopes, corresponding to our total measured range of [Al/Fe]$\sim0.2$\,dex, the models of \citet{denissenkov} indeed suggest an insignificant variation in the \mgi\ and \mgiii\ fractions ($<5$\%), as less than 10\% of the supermassive star ejecta will need to be incorporated in the material  making up the polluted population. On the other hand, taking into account the additional constraints from the Na-O anti-correlation, this changes the picture. If the yields are shifted by $+0.1$\,dex in [O/Fe] to match our most oxygen-rich star (excluding star 10237), 30-70\% material from the supermassive stars is required to explain the variation in our observed sample. This would imply a variation of [Al/Fe] of about 1.2 dex, and a depletion of \mgi\ by up to 35\%, with an increase of \mgiii\ by an equivalent amount, relative to the pristine mixture. This is in stark contrast to what we observe, and thus supermassive stars are not able to explain the full set of abundance variations in the case of 47 Tucanae. The large amount of polluted material required to explain the Na-O anti-correlation would also imply a broad He variation ($\Delta Y\sim0.1$). This is in contrast to the observed spread in He of $\Delta Y=\sim0.03$ \citep{dicriscienzo}.

However, if one considers only the variations in [Al/Fe] and in the heavy Mg isotopes, a significant variation in He may not be required. This is indeed consistent with the observations here, but would then require a separate mechanism for explaining the Na-O variations, in particular a mechanism that does not simultaneously produce large amounts of He.

Whereas the AGB star and supermassive star scenarios are the only ones able to \emph{produce} the heavy Mg isotopes, other mechanisms could explain the observed Na-O anti-correlation. In addition, some alternative models may be able to modify \mgi, which would also change the isotopic ratios. Unfortunately, the lack of models at metallicities appropriate for 47Tuc means that it is not possible to make as detailed a comparison as for the AGB stars. 

\citet{decressin} proposes fast rotating, massive stars ($20$\msun\ $<M<120$\msun, FRMS) as polluter candidates. Here, the central idea is that the abundance variations are created during hydrostatic burning and subsequently transported to the surface through rotationally-induced mixing. Here, the gas is ejected from the star through a slow wind, so that it is possible to retain the enriched gas within the cluster. The models considered in their work ([Fe/H]$\approx-1.5$), exhibit a strong enhancement in Na at the stellar surface, together with depletion in O, when their models reach the end of the main sequence. The stellar wind is slow at this evolutionary stage, and thus has a composition appropriate for the polluted population of stars in GCs. At the metallicity considered in their study, the increase in Na is between 0.8 and 1.6 dex, depending on the adopted reaction rates, while O is depleted by about 1.0 dex, compared to their initial values. This is expected to hold also at higher metallicities, where the core temperatures tend to increase, due to the increased opacity. This should result in an even more efficient burning. Even just assuming that the yields stay the same at the metallicity of 47Tuc, they can easily accommodate both the 0.48 dex depletion we find in [O/Fe], and the associated 0.49 dex increase in [Na/Fe]. However, dedicated models at higher metallicities are much called for, as the yields are complex functions of nuclear burning, stellar evolution and mass-loss, all of which are influenced by metallicity to some extent. As mentioned earlier, these models result in a net production of Mg, largely in the form of \mgi, so one would expect the isotopic ratios to decrease with increasing Al enhancement, clearly contradicting observations. 

The scenario proposed by \citet{demink}, hereafter dM09, is exploring the viability of massive interacting binaries as the source of the enrichment. This polluter candidate has some appealing properties, compared to the AGB and FRMS scenarios. In particular, it provides a very efficient way of releasing large amounts of enriched material into the cluster environment through mass transfer, whereas the two other scenarios require the cluster to either have been significantly more massive in the past, or have a very peculiar initial mass function. Unfortunately, this candidate has seen very little additional research, and dM09 only consider the single case of a 12\msun\ and 20\msun\ system at a metallicity of [Fe/H] $\approx-1.5$. Their yields are similar to what is found by \citet{decressin}, with their average yields showing a 0.12 dex depletion of oxygen, a 1.0 dex increase in sodium and a 0.13 dex increase in Al. Interestingly, the increase in Al comes from processing of Mg, which is slightly depleted. Whereas dM09 do not provide yields for the individual isotopes of Mg, this could result in an increase in the \mgii/\mgi\ and \mgiii/\mgi\ ratios, if it is assumed that the Al production comes mainly from burning of \mgi. 

Their most extreme values for abundance variations are somewhat higher, and can easily accommodate the range we observe, under the assumption of identical yields at higher [Fe/H]. Furthermore, since the mass transfer between the massive binaries will spin up the companion star \citep{demink2}, rotationally induced mixing may alter the abundances of the distributed material further. But as with the FRMS scenario, a larger suite of models would be most welcome to investigate this proposed mechanism in more detail. We note here that the pollution scenario using early disk accretion \citep{bastian} uses the same interacting binaries as the main polluter source, so the chemical abundance pattern in from the accretion mechanism will be similar to what is provided by the interacting binaries. 

\subsection{Isotopes in 3D. A partial solution to the \mgii\ problem}
The discrepancy between the observed values for \mgii, and what is predicted to be produced by AGB stars, has been a long-standing issue, and the essentially constant, low fraction of \mgii\ is seen in all GCs where this has been studied, as mentioned earlier. Despite the improvements in AGB nucleosynthesis over the last decade, the predicted isotopic ratios are still at odds with the observations. \citet{dacosta} propose two different scenarios that could potentially explain the observed behaviour. Both scenarios require modification of nuclear reaction rates at levels that are essentially ruled out by nuclear physics.

\begin{figure}%
\centering
\includegraphics[trim= 6cm 5cm 0cm 0cm, width=\columnwidth]{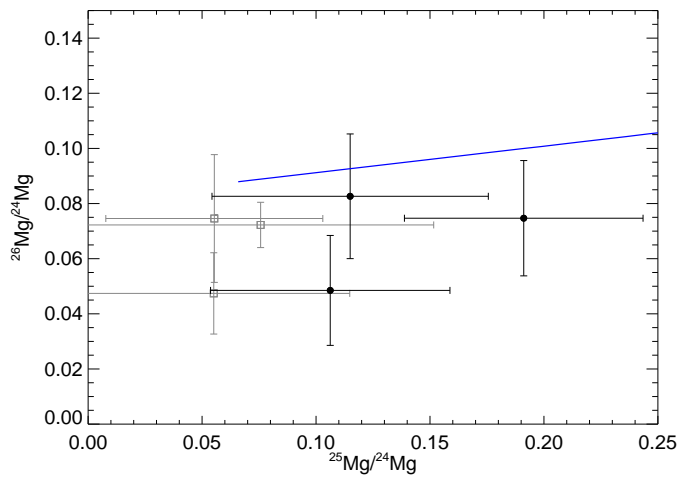}%
\caption{Isotopic ratios of magnesium from 1D (gray, open squares) and 3D (black, filled circles), together with the dilution curve for AGB star ejecta. Results are shown for stars 4794, 13396 and 29861.}%
\label{dilution3D}%
\end{figure}

Our 3D results may resolve part of this discrepancy, as can be seen by inspecting Fig~\ref{dilution3D}. Here we show the mean values for the isotopic ratios for the three stars for which we have performed a full 3D synthesis. For comparison, we also show the 1D results. We note that these mean values differ somewhat from what was shown in Fig~\ref{ratios}, since we only use the 5135\,\AA\ and 5138\,\AA\ features. This is also the explanation for the increased uncertainties which we give as the RMS error of the mean value. For star 29861, where we only have results from one feature, we use the mean of the uncertainties of the two others stars as an estimate of the typical scatter. 

The effect of the 3D synthesis is evident. We observe a factor of $2-2.5$ increase in the \mgii/\mgi\ ratios, with respect to 1D, whereas the \mgiii/\mgi\ values stay essentially unchanged, compared to 1D. The measurements now show a significantly better agreement with the predictions from the AGB models. The \mgii/\mgi\ ratios are now well within the ratios predicted for almost-pristine composition stars in 47Tuc, whereas there still seems to be a small offset in \mgiii/\mgi, compared to the models. This may potentially be rectified if results from the 5134\,\AA\ and 5140\,\AA\ features are included, as they tend to yield higher \mgiii\ fractions, compared to the two bands investigated here. It is also worth noting that in all cases, the amount of \mgii\ is now \textit{above} that of \mgiii, which is in much better agreement with the predictions from AGB stars.  

\begin{figure}%
\centering
\includegraphics[trim = 6cm 8cm 0cm 0cm, width=\columnwidth]{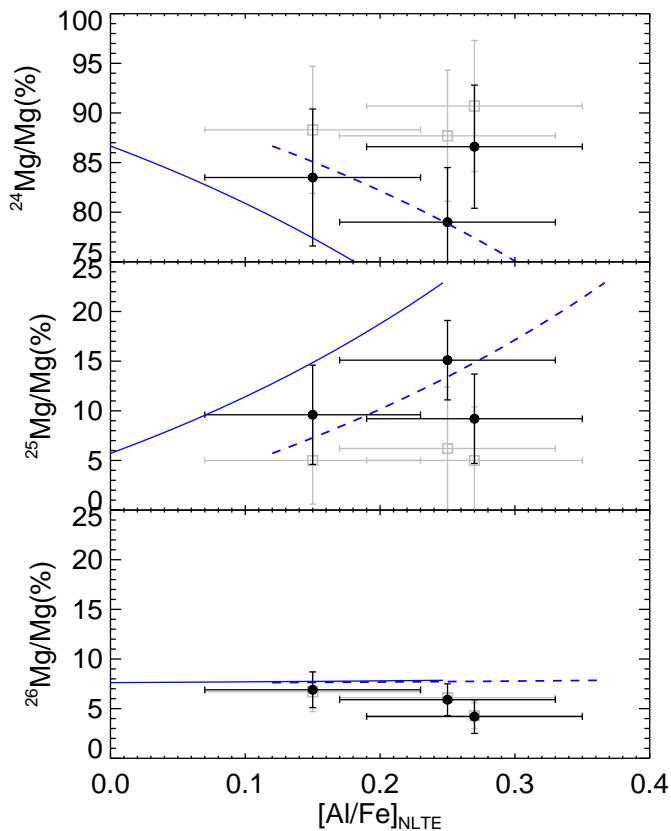}%
\caption{Isotopic fractions vs. [Al/Fe] for the stars with 3D results. The solid blue lines is the predicted composition of the stars from \citet{ventura}. The dashed lines shows the dilution curve, when shifted by +0.12 dex in [Al/Fe]. Symbols have the same meaning as in Fig.~\ref{dilution3D}.}%
\label{iso-al-3D}%
\end{figure}

That the effects of 3D atmospheres are indeed improving the agreement with the AGB models can also be seen in Fig.\ref{iso-al-3D}, where we again plot the isotopic fractions vs. [Al/Fe], as in Fig.~\ref{iso-al}. The two dilution curves have the same meaning as before. The \mgi\ and \mgii\ fractions now fall very close to the dilution curve, if we apply the +0.12 dex shift to the [Al/Fe] predictions from the models. This shows that if the AGB models start out with a slightly different initial composition, it may indeed be possible to get a good agreement between models and observations of 47Tuc.

Whereas these results are encouraging, and may hopefully resolve a large part of the discrepancy between the observations and the predictions for AGB polluters in GCs, it may introduce another problem. If this result holds in general, an increase of \mgii\ would be expected also for giants in the field. Here, the predictions from chemical evolution models suggest that the models under-produce the amount of heavy isotopes \citep{kobayashi}, and our results indicate that this discrepancy will increase in magnitude, if the field stars are investigated with 3D atmospheres. On the other hand, most of the stars used in their study are dwarfs, and the impact of using 3D atmospheres in dwarf stars will likely be different and may mitigate part of this discrepancy.

We caution that a larger grid of 3D stellar atmosphere models is needed before the full impact of 3D synthesis can be determined. Whereas our syntheses are interpolated to the observed metallicities, we only cover a single value of \teff\ and \logg. It would be desirable to cover also these dimensions in the parameter space, to allow us to interpolate to the exact stellar parameters, rather than our current approach. This may change the observed ratios somewhat, but we consider our model parameters to be close enough to the actual stellar parameters, to give at least a qualitative indication of the expected 3D effects for giants at these parameters. Efforts are currently ongoing to expand the available parameter space of our 3D models, so that these issues can be addressed in more detail. This also holds for analysis of stars in the field, where a larger range in model metallicities will also be required.

These results do also not provide an answer to the discrepancy between the observed and predicted behaviour of the \textit{polluted} population of stars. Here, the AGB models predict a significant increase in the \mgii/\mgi\ ratio and this ratio should be higher than the \mgiii/\mgi\ ratio for the same stars. Since the stars for which we investigate the 3D effects, all belong to the pristine population, the results shown here cannot be used to determine any differences between the two populations. So while 3D effects are likely to explain a large part of the apparent lack of \mgii\ previously reported, they are unlikely to resolve the problem of the lack of variation between the populations. Since our polluted population of stars all have significantly lower \logg\ values, we cannot rule out a possible differential 3D effect, since the convective broadening of spectral lines tend to increase with decreasing surface gravity. However, we have no a priori reason to believe that the changes in the isotopic composition due to 3D effects should be significantly different for the polluted population of stars, compared to the pristine population, for similar stellar parameters. 

\section{Conclusions}
In this paper we have presented the first ever measurements of the Mg isotopic distribution in giants in the globular cluster 47 Tucanae. The sample shows no significant variation in the isotopic fractions, but given the negligible variation in Al for our sample of stars, this is not unexpected. In lower-metallicity clusters like NGC6752 \citep{yong6752}, where a stronger Al enhancement is observed, the isotope variations are always found to be more substantial. We were not able to detect any significant correlation between the isotopes and [Na/Fe] or [Al/Fe], which has been observed in clusters with broader Al variations. As such, we do not see any evidence for any significant activation of the Mg-Al burning chain. We attribute this to our small sample size, since we do not sample the most Al-enriched stars in this cluster, as seen by other authors  (\citealt{carretta2013}; \citealt{cordero}). These stars are expected to coincide with the stars most strongly enhanced in the heavy Mg isotopes, so the true variation in Mg isotopes is potentially broader than what presented here.

In addition, we provide the first, detailed investigation of Mg isotopes with the use of 3D hydrodynamical atmospheres and full 3D spectral synthesis. The 3D synthesis provides an improved fit to the observed features, with significant changes in the \mgii/\mgi\ ratio, by up to a factor of 2.5. This isotopic ratio is found to increase in both MgH features investigated. The \mgiii/\mgi\ ratio, on the other hand, is essentially unchanged. A particularly interesting aspect is that the fraction of \mgii\ is now found to be higher than that of \mgiii, which has not been observed before. This helps to resolve a large part of the discrepancy between the AGB model yields and observations for this cluster, where the observed amounts of \mgii, based on 1D model atmospheres, are significantly below what is predicted. In particular if it is assumed that the most [Al/Fe] poor star in our sample represents the pristine composition. However, a larger sample of stars is needed to establish this firmly. Whereas the 3D results are encouraging, we caution that the use of 3D syntheses are unlikely to resolve the discrepancy between the observed and predicted variations of Mg isotopes between the pristine and polluted population under the AGB scheme. The increase in the \mgii/\mgi\ ratio would be expected to be similar in field stars, and this may result in increased tension between the predictions from chemical evolution models of the Milky Way, and the observed abundance ratios.

The main reason for the different results between 1D and 3D was found to be related to the temperature fluctuations in the 3D models, since the average thermal structures are almost identical at these metallicities. We caution that since we do not investigate 3D effects for all four MgH features used, it is premature to draw conclusions on the overall effect of 3D on the isotopic abundances. This will require a larger model grid as well as a much more detailed investigation, which is beyond the scope of this paper. The results are encouraging and certainly warrant more detailed research into the effects of 3D atmospheres on Mg isotopes, for a broader range of parameters and metallicities. In particular it would be interesting to also investigate this for field stars in the Galactic halo, where the Mg isotopic ratios are predominantly derived from sub-dwarfs (52/61 stars \citealt{yong_field}), where the effects of 3D atmospheres will likely be different. On the other hand, the nine giants also investigated in their work, do have isotopic ratios similar to the dwarfs at the same metallicity.

It would be desirable to obtain new observations at the extremes of the [Al/Fe] variations, to measure the true extent of the Mg isotope variations. This could provide additional support for massive AGB stars having contributed significantly to the intra-cluster pollution of 47Tuc at early times. 

\begin{acknowledgements} 
The authors would like to thank the anonymous referee for their useful comments which helped improve the manuscript. AOT, LS, HGL and NC acknowledges support from Sonderforschungsbereich SFB 881 "The Milky Way System" (subprojects A4 and A5) of the German Research Foundation (DFG). LS acknowledges the support by Chile's Ministry of Economy, Development, and Tourism's Millennium Science Initiative through grant IC120009, awarded to The Millennium Institute of Astrophysics, MAS. DY was supported through an Australian Research Council Future Fellowship (FT140100554). RC acknowledges support from a DECRA grant from the Australian Research Council (project DE120102940). This work has made use of the VALD database, operated at Uppsala University, the Institute of Astronomy RAS in Moscow, and the University of Vienna. This research took advantage of the SIMBAD and VIZIER databases at the CDS, Strasbourg (France), and NASA's Astrophysics Data System Bibliographic Services. 
\end{acknowledgements}

\bibliographystyle{aa}
\bibliography{47Tuc_paper2} 

\appendix

\end{document}

%% file: fitprofile.tex
\newcommand{\fitp}{\texttt{Fitprofile}}
\newcommand{\mygi}{MyGIsFOS} 

The software used for the 1D fitting of the MgH features using MOOG calculates spectral synthesis on the fly, which is not feasible to do in 3D, where the computational overheads are significantly higher. Thus, for the purpose of determining the best-fitting 3D synthetic profile we developed the multi-parametric fitting code \fitp, which uses a pre-computed grid of syntheses. Before applying to our 3D synthesis, we tested that the 1D synthesis provided near-identical results to our 1D fitting routines using MOOG.

\fitp\ shares much of the general inner workings and a relevant amount of code with the automated parameter determination and abundance analysis code \mygi\ \citep{sbordone14}. As \mygi\, it is written in Fortran 90 (with Intel extensions) and uses the CERN function minimisation library MINUIT \citep{minuit1,minuit2} as $\chi^2$ minimisation engine. The \fitp\ code was developed to provide a flexible line-fitting tool that could both perform simple, general purpose line fitting for single lines, and more sophisticated multi-parameter, multi-region fitting such as the one needed for the present study. \fitp\ was used already in Paper 1, to derive abundances from elements with lines that showed hyperfine splittings. 

\fitp\ reads in a grid of synthetic spectra varying in up to four parameters, an observed spectrum against which the fit is to be performed, and a list of spectral regions to be used in the fit. Two types of regions are accepted: pseudo-continuum regions are used to pseudo-normalise the observed spectrum as well as the synthetic grid, and to determine the S/N ratio, while fitting regions define the spectral ranges over which the actual fitting is performed. We used the same fitting regions for each feature as in the 1D case.

The up-to-four parameters varying in the grid can represent any quantity varying from one spectrum to another in the synthetic grid. Any number of them can be fitted, or kept fixed to an user-chosen value. In the case of the present work, for instance, 
the provided grid parameters were set to total Mg abundance, $^{25}$Mg fraction and $^{26}$Mg fraction (and all were fitted), while the fourth parameter was not used. From this, the \mgi\ fraction could easily be calculated. The grid must be equispaced and rectangular in any parameter it contains. 

If the user provides pseudo-normalisation regions, they are used to locally estimate the S/N and a pseudo-continuum spline that is used to pseudo-normalise the synthetic grid and observed spectrum. Both quantities are estimated as constant if one single pseudo-continuum interval was provided, as a linear interpolation if two pseudo-continua are given, and as a spline for three or more pseudo-continua. Both quantities can be kept fixed, providing a pre-normalised spectrum and an estimation of S/N. Due to the lack of suitable continuum regions in the immediate vicinity of the MgH bands, we used pre-normalised spectra with a S/N estimate for our fitting. The normalisation was identical to what was used in MOOG.

Fitting regions can be contiguous or not. A contiguous region is, for instance, a typical spectral region covering a spectral line 
one wants to fit. A non-contiguous region is called in \fitp\ a ``region group'', as it is, indeed, a group of contiguous regions
that are fitted as one, i.e. a single $\chi^2$ is computed for all the pixels contained in the contiguous regions composing the group. \fitp\ produces fit values for each provided region group. To apply this to the present work, one could have fitted all the MgH features together, deriving a single best fitting set of Mg isotopic ratios, by including them all in a single group. Or, as we did, fit each feature independently, producing multiple sets of best fitting values that can then be averaged. In fact, within \fitp\ every contiguous feature fitted on its own is described as a one-feature group.

During the fitting, two additional parameters can be allowed to vary, namely some adjustment of the continuum value (up to a fraction of the local S/N), and some amount of Doppler shift (within user-defined limits). These adjustments are applied to the pre-computed syntheses grid and applied per feature group. Both parameters can be be disabled if the user so desire. In this case, we allowed for small continuum adjustments as well as a small velocity shift, both of which are included in the fitting. Upon a successful fit, \fitp\ provides the best fitting values of the grid parameters for each group as well as an average among the groups and fitted profiles for each group for inspection of the results. For each individual group, \fitp\ also provides the best-fitting doppler shift.

Another useful capability of \fitp\ is to allow parameter space mapping, i.e., aside from the MINUIT $\chi^2$ minimum search, the $\chi^2$ value is computed at each grid point, allowing the uncertainty of the fit, and the parameter correlation to be estimated.

%% file: mg-isotopes3.tex
ID & 5134.57\AA & 5135.07\AA & 5138.71\AA & 5140.20\AA & Mean & $\sigma_{tot.}$\\
\hline\hline 
 4794 & 85.3:3.2:11.5 & 82.0:9.4:8.6 & 94.7:0.6:4.7 & 89.5:0.9:9.5 & 88.1:4.0:7.9 & 2.9:2.1:1.6 \\
 5968 & 84.9:3.4:11.7 & 85.0:8.1:6.9 & 95.2:1.2:3.6 & 89.8:3.6:6.6 & 89.1:4.3:6.6 & 2.6:1.5:1.8 \\
 6798 & 86.3:3.6:10.1  & 85.5:8.4:6.1 & 95.6:0.8:3.6 & 90.3:4.7:5.0 & 89.7:4.5:5.8 & 2.5:1.7:1.6 \\
10237 & 75.4:13.0:11.6 & 72.5:18.4:9.1 & 88.5:5.3:6.2 & 87.8:6.4:5.8 & 80.9:11.1:8.0 & 4.3:3.1:1.5 \\
13396 & 84.4:5.6:10.0 & 81.2:12.3:6.5 & 94.3:0.0:5.7 & 88.9:0.9:10.2 & 87.4:5.2:7.4 & 3.0:2.9:1.4 \\
20885 & 80.6:7.4:12.0 & Discarded & Discarded & 91.2:0.4:8.4 & 85.9:3.9:10.2 & 5.4:3.5:1.9 \\
29861 & 74.2:13.1:12.7 & Discarded & 90.7:5.0:4.3 & 88.4:2.7:8.9 & 85.9:6.5:7.6 & 5.1:2.9:2.7 \\
\textbf{1062} & 84.0:2.7:13.3 & 88.7:5.3:6.0 & 95.8:0.9:3.3 & 91.1:1.0:7.9 & 90.7:2.7:6.6 & 2.7:1.1:2.2 \\
\textbf{5265} & 83.9:3.5:12.6 & 86.8:4.2:9.0 & 95.6:0.5:3.9 & 85.7:5.8:8.5 & 89.0:3.1:7.8 & 2.8:1.2:1.9 \\
\textbf{27678} & 81.4:1.9:16.7 & 86.6:4.2:9.2 & 91.9:2.2:5.9 & 87.5:2.1:10.4 & 87.6:2.8:9.6 & 2.4:0.7:2.4 \\
\textbf{28956} & Discarded & 82.8:5.4:11.8 & 93.4:0.7:5.9 & 82.0:6.2:11.8 & 86.9:3.7:9.4 & 4.0:1.8:2.2 \\
\textbf{38916} & 80.4:5.9:13.7 & 79.4:11.2:9.4  & 92.4:1.6:6.0 & 81.0:9.4:9.6 & 84.2:6.8:9.0 & 3.2:2.2:1.7 \\
\textbf{40394} & 85.7:1.4:12.9 & 91.7:2.4:5.9 & 98.1:0.1:1.8 & 90.6:3.1:6.3 & 92.7:1.6:5.7 & 2.8:0.8:2.4 \\
\hline

%% file: mg-isotopes-avg3.tex
ID & \teff & log(g) & $\xi_t$ & [Fe/H] & [C/Fe] & [O/Fe] & [Na/Fe]$_\mathrm{NLTE}$ & [Mg/Fe] & [Al/Fe]$_\mathrm{NLTE}$ & $^{24}$Mg:$^{25}$Mg:$^{26}$Mg \\
\hline\hline
 4794 & 4070 & 1.15 & 1.30 & $-0.66$ & $-0.16$ & 0.38 & 0.15 & 0.40 & 0.15 &  88.1:4.0:7.9 \\
 5968 & 3970 & 0.85 & 1.40 & $-0.79$ & $+0.01$ & 0.41 & 0.08 & 0.49 & 0.17 &  89.1:4.3:6.6 \\
 6798 & 4000 & 0.90 & 1.30 & $-0.69$ & $-0.04$ & 0.37 & 0.01 & 0.44 & 0.12 &  89.7:4.5:5.8 \\
10237 & 4280 & 1.20 & 1.60 & $-0.83$ & $+0.01$ & 0.57 & 0.04 & 0.32 & 0.18 &  80.9:11.1:8.0 \\
13396 & 4190 & 1.45 & 1.60 & $-0.83$ & $-0.09$ & 0.48 & 0.07 & 0.44 & 0.25 &  87.4:5.2:7.4 \\
20885 & 4260 & 1.35 & 1.90 & $-0.84$ & $-0.12$ & $-$ & 0.11 & 0.47 & 0.23 &  85.9:3.9:10.2 \\
29861 & 4160 & 1.20 & 1.50 & $-0.84$ & $+0.20$ & 0.47 & 0.10 & 0.40 & 0.27 &  85.9:6.5:7.6 \\
\textbf{1062} & 3870 & 0.45 & 1.30 & $-0.78$ & $-0.34$ & 0.23 & 0.24 & 0.46 & 0.16 &  90.7:2.7:6.6 \\
\textbf{5265} & 3870 & 0.30 & 1.25 & $-0.69$ & $-0.31$ & 0.00 & 0.31 & 0.40 & 0.19 &  89.0:3.1:7.8 \\
\textbf{27678} & 3870 & 0.35 & 1.20 & $-0.76$ & $-0.48$ & 0.13 & 0.50 & 0.45 & 0.29 &  87.6:2.8:9.6 \\
\textbf{28956} & 3900 & 0.30 & 1.60 & $-0.86$ & $-0.24$ & 0.06 & 0.41 & 0.46 & 0.29 &  86.9:3.7:9.4 \\
\textbf{38916} & 4080 & 0.85 & 1.40 & $-0.83$ & $-0.42$ & $-$ & 0.42 & 0.52 & 0.29 &  84.2:6.8:9.0 \\
\textbf{40394} & 3890 & 0.45 & 1.10 & $-0.71$ & $-0.32$ & 0.23 & 0.26 & 0.43 & 0.12 &  92.7:1.6:5.7 \\
\hline

%% file: 47Tuc_paper2_v8.bbl
\begin{thebibliography}{77}
\expandafter\ifx\csname natexlab\endcsname\relax\def\natexlab#1{#1}\fi

\bibitem[{{Alib{\'e}s} {et~al.}(2001){Alib{\'e}s}, {Labay}, \&
  {Canal}}]{alibes}
{Alib{\'e}s}, A., {Labay}, J., \& {Canal}, R. 2001, \aap, 370, 1103

\bibitem[{{Alves-Brito} {et~al.}(2005){Alves-Brito}, {Barbuy}, {Ortolani},
  {Momany}, {Hill}, {Zoccali}, {Renzini}, {Minniti}, {Pasquini}, {Bica}, \&
  {Rich}}]{alves-brito}
{Alves-Brito}, A., {Barbuy}, B., {Ortolani}, S., {et~al.} 2005, \aap, 435, 657

\bibitem[{{Anderson} {et~al.}(2009){Anderson}, {Piotto}, {King}, {Bedin}, \&
  {Guhathakurta}}]{anderson}
{Anderson}, J., {Piotto}, G., {King}, I.~R., {Bedin}, L.~R., \& {Guhathakurta},
  P. 2009, \apjl, 697, L58

\bibitem[{{Asplund} {et~al.}(2009){Asplund}, {Grevesse}, {Sauval}, \&
  {Scott}}]{asplund}
{Asplund}, M., {Grevesse}, N., {Sauval}, A.~J., \& {Scott}, P. 2009, \araa, 47,
  481

\bibitem[{{Bastian} {et~al.}(2015){Bastian}, {Cabrera-Ziri}, \&
  {Salaris}}]{bastian2}
{Bastian}, N., {Cabrera-Ziri}, I., \& {Salaris}, M. 2015, \mnras, 449, 3333

\bibitem[{{Bastian} {et~al.}(2013){Bastian}, {Lamers}, {de Mink}, {Longmore},
  {Goodwin}, \& {Gieles}}]{bastian}
{Bastian}, N., {Lamers}, H.~J.~G.~L.~M., {de Mink}, S.~E., {et~al.} 2013,
  \mnras, 436, 2398

\bibitem[{{Bernath} {et~al.}(1985){Bernath}, {Black}, \& {Brault}}]{bernath}
{Bernath}, P.~F., {Black}, J.~H., \& {Brault}, J.~W. 1985, \apj, 298, 375

\bibitem[{{Briley}(1997)}]{1997AJ....114.1051B}
{Briley}, M.~M. 1997, \aj, 114, 1051

\bibitem[{{Briley} {et~al.}(2001){Briley}, {Smith}, \&
  {Claver}}]{2001AJ....122.2561B}
{Briley}, M.~M., {Smith}, G.~H., \& {Claver}, C.~F. 2001, \aj, 122, 2561

\bibitem[{{Briley} {et~al.}(1996){Briley}, {Smith}, {Suntzeff}, {Lambert},
  {Bell}, \& {Hesser}}]{briley2}
{Briley}, M.~M., {Smith}, V.~V., {Suntzeff}, N.~B., {et~al.} 1996, \nat, 383,
  604

\bibitem[{{Brooke} {et~al.}(2013){Brooke}, {Bernath}, {Schmidt}, \&
  {Bacskay}}]{brooke}
{Brooke}, J.~S.~A., {Bernath}, P.~F., {Schmidt}, T.~W., \& {Bacskay}, G.~B.
  2013, \jqsrt, 124, 11

\bibitem[{{Brooke} {et~al.}(2014){Brooke}, {Ram}, {Western}, {Li}, {Schwenke},
  \& {Bernath}}]{2014ApJS..210...23B}
{Brooke}, J.~S.~A., {Ram}, R.~S., {Western}, C.~M., {et~al.} 2014, \apjs, 210,
  23

\bibitem[{{Caffau} \& {Ludwig}(2007)}]{caffau2007}
{Caffau}, E. \& {Ludwig}, H.-G. 2007, \aap, 467, L11

\bibitem[{{Carretta} {et~al.}(2009){Carretta}, {Bragaglia}, {Gratton}, \&
  {Lucatello}}]{carretta2009}
{Carretta}, E., {Bragaglia}, A., {Gratton}, R., \& {Lucatello}, S. 2009, \aap,
  505, 139

\bibitem[{{Carretta} {et~al.}(2013){Carretta}, {Gratton}, {Bragaglia},
  {D'Orazi}, \& {Lucatello}}]{carretta2013}
{Carretta}, E., {Gratton}, R.~G., {Bragaglia}, A., {D'Orazi}, V., \&
  {Lucatello}, S. 2013, \aap, 550, A34

\bibitem[{{Carretta} {et~al.}(2005){Carretta}, {Gratton}, {Lucatello},
  {Bragaglia}, \& {Bonifacio}}]{carretta}
{Carretta}, E., {Gratton}, R.~G., {Lucatello}, S., {Bragaglia}, A., \&
  {Bonifacio}, P. 2005, \aap, 433, 597

\bibitem[{{Collet} {et~al.}(2007){Collet}, {Asplund}, \& {Trampedach}}]{collet}
{Collet}, R., {Asplund}, M., \& {Trampedach}, R. 2007, \aap, 469, 687

\bibitem[{{Cordero} {et~al.}(2014){Cordero}, {Pilachowski}, {Johnson},
  {McDonald}, {Zijlstra}, \& {Simmerer}}]{cordero}
{Cordero}, M.~J., {Pilachowski}, C.~A., {Johnson}, C.~I., {et~al.} 2014, \apj,
  780, 94

\bibitem[{{Cottrell} \& {Da Costa}(1981)}]{cottrell}
{Cottrell}, P.~L. \& {Da Costa}, G.~S. 1981, \apjl, 245, L79

\bibitem[{{Da Costa} {et~al.}(2013){Da Costa}, {Norris}, \& {Yong}}]{dacosta}
{Da Costa}, G.~S., {Norris}, J.~E., \& {Yong}, D. 2013, \apj, 769, 8

\bibitem[{{Dalessandro} {et~al.}(2013){Dalessandro}, {Salaris}, {Ferraro},
  {Mucciarelli}, \& {Cassisi}}]{2013MNRAS.430..459D}
{Dalessandro}, E., {Salaris}, M., {Ferraro}, F.~R., {Mucciarelli}, A., \&
  {Cassisi}, S. 2013, \mnras, 430, 459

\bibitem[{{D'Antona} \& {Ventura}(2007)}]{dantona_mixing}
{D'Antona}, F. \& {Ventura}, P. 2007, \mnras, 379, 1431

\bibitem[{{de Mink} {et~al.}(2013){de Mink}, {Langer}, {Izzard}, {Sana}, \& {de
  Koter}}]{demink2}
{de Mink}, S.~E., {Langer}, N., {Izzard}, R.~G., {Sana}, H., \& {de Koter}, A.
  2013, \apj, 764, 166

\bibitem[{{de Mink} {et~al.}(2009){de Mink}, {Pols}, {Langer}, \&
  {Izzard}}]{demink}
{de Mink}, S.~E., {Pols}, O.~R., {Langer}, N., \& {Izzard}, R.~G. 2009, \aap,
  507, L1

\bibitem[{{Decressin} {et~al.}(2007){Decressin}, {Meynet}, {Charbonnel},
  {Prantzos}, \& {Ekstr{\"o}m}}]{decressin}
{Decressin}, T., {Meynet}, G., {Charbonnel}, C., {Prantzos}, N., \&
  {Ekstr{\"o}m}, S. 2007, \aap, 464, 1029

\bibitem[{{Dekker} {et~al.}(2000){Dekker}, {D'Odorico}, {Kaufer}, {Delabre}, \&
  {Kotzlowski}}]{dekker}
{Dekker}, H., {D'Odorico}, S., {Kaufer}, A., {Delabre}, B., \& {Kotzlowski}, H.
  2000, in SPIE Conference Series, Vol. 4008, SPIE Conference Series, ed.
  M.~{Iye} \& A.~F. {Moorwood}, 534--545

\bibitem[{{Denissenkov} \& {Hartwick}(2014)}]{denissenkov}
{Denissenkov}, P.~A. \& {Hartwick}, F.~D.~A. 2014, \mnras, 437, L21

\bibitem[{{Denissenkov} {et~al.}(2015){Denissenkov}, {VandenBerg}, {Hartwick},
  {Herwig}, {Weiss}, \& {Paxton}}]{2015MNRAS.448.3314D}
{Denissenkov}, P.~A., {VandenBerg}, D.~A., {Hartwick}, F.~D.~A., {et~al.} 2015,
  \mnras, 448, 3314

\bibitem[{{Denissenkov} {et~al.}(1997){Denissenkov}, {Weiss}, \&
  {Wagenhuber}}]{denissenkov97}
{Denissenkov}, P.~A., {Weiss}, A., \& {Wagenhuber}, J. 1997, \aap, 320, 115

\bibitem[{{di Criscienzo} {et~al.}(2010){di Criscienzo}, {Ventura}, {D'Antona},
  {Milone}, \& {Piotto}}]{dicriscienzo}
{di Criscienzo}, M., {Ventura}, P., {D'Antona}, F., {Milone}, A., \& {Piotto},
  G. 2010, \mnras, 408, 999

\bibitem[{{Dravins}(1982)}]{dravins}
{Dravins}, D. 1982, \araa, 20, 61

\bibitem[{{Fishlock} {et~al.}(2014){Fishlock}, {Karakas}, {Lugaro}, \&
  {Yong}}]{fishlock}
{Fishlock}, C.~K., {Karakas}, A.~I., {Lugaro}, M., \& {Yong}, D. 2014, \apj,
  797, 44

\bibitem[{{Freytag} {et~al.}(2012){Freytag}, {Steffen}, {Ludwig},
  {Wedemeyer-B{\"o}hm}, {Schaffenberger}, \& {Steiner}}]{co5bold}
{Freytag}, B., {Steffen}, M., {Ludwig}, H.-G., {et~al.} 2012, Journal of
  Computational Physics, 231, 919

\bibitem[{{Gratton} {et~al.}(2012){Gratton}, {Carretta}, \&
  {Bragaglia}}]{gratton}
{Gratton}, R.~G., {Carretta}, E., \& {Bragaglia}, A. 2012, \aapr, 20, 50

\bibitem[{{Hinkle} {et~al.}(2013){Hinkle}, {Wallace}, {Ram}, {Bernath},
  {Sneden}, \& {Lucatello}}]{hinkle}
{Hinkle}, K.~H., {Wallace}, L., {Ram}, R.~S., {et~al.} 2013, \apjs, 207, 26

\bibitem[{{James} \& {Roos}(1975)}]{minuit1}
{James}, F. \& {Roos}, M. 1975, Computer Physics Communications, 10, 343

\bibitem[{{Johnson} {et~al.}(2014){Johnson}, {Rich}, {Kobayashi}, {Kunder}, \&
  {Koch}}]{2014AJ....148...67J}
{Johnson}, C.~I., {Rich}, R.~M., {Kobayashi}, C., {Kunder}, A., \& {Koch}, A.
  2014, \aj, 148, 67

\bibitem[{{Karakas}(2010)}]{karakas2}
{Karakas}, A.~I. 2010, in IAU Symposium, Vol. 266, IAU Symposium, ed. R.~{de
  Grijs} \& J.~R.~D. {L{\'e}pine}, 161--168

\bibitem[{{Karakas} \& {Lattanzio}(2003)}]{karakas}
{Karakas}, A.~I. \& {Lattanzio}, J.~C. 2003, \pasa, 20, 279

\bibitem[{{Karakas} {et~al.}(2006){Karakas}, {Lugaro}, {Wiescher},
  {G{\"o}rres}, \& {Ugalde}}]{karakas2006}
{Karakas}, A.~I., {Lugaro}, M.~A., {Wiescher}, M., {G{\"o}rres}, J., \&
  {Ugalde}, C. 2006, \apj, 643, 471

\bibitem[{{Kobayashi} {et~al.}(2011){Kobayashi}, {Karakas}, \&
  {Umeda}}]{kobayashi}
{Kobayashi}, C., {Karakas}, A.~I., \& {Umeda}, H. 2011, \mnras, 414, 3231

\bibitem[{{Koch} \& {McWilliam}(2008)}]{koch}
{Koch}, A. \& {McWilliam}, A. 2008, \aj, 135, 1551

\bibitem[{{Kupka} {et~al.}(2000){Kupka}, {Ryabchikova}, {Piskunov}, {Stempels},
  \& {Weiss}}]{vald}
{Kupka}, F.~G., {Ryabchikova}, T.~A., {Piskunov}, N.~E., {Stempels}, H.~C., \&
  {Weiss}, W.~W. 2000, Baltic Astronomy, 9, 590

\bibitem[{{Lazzaro} \& {Moneta}(2010)}]{minuit2}
{Lazzaro}, A. \& {Moneta}, L. 2010, Journal of Physics Conference Series, 219,
  042044

\bibitem[{Longland {et~al.}(2010)Longland, Iliadis, Champagne, Newton, Ugalde,
  Coc, \& Fitzgerald}]{Longland20101}
Longland, R., Iliadis, C., Champagne, A., {et~al.} 2010, Nuclear Physics A,
  841, 1 , the 2010 Evaluation of Monte Carlo based Thermonuclear Reaction
  Rates

\bibitem[{{Markwardt}(2009)}]{markwardt}
{Markwardt}, C.~B. 2009, in Astronomical Society of the Pacific Conference
  Series, Vol. 411, Astronomical Data Analysis Software and Systems XVIII, ed.
  D.~A. {Bohlender}, D.~{Durand}, \& P.~{Dowler}, 251

\bibitem[{{Masseron} {et~al.}(2014){Masseron}, {Plez}, {Van Eck}, {Colin},
  {Daoutidis}, {Godefroid}, {Coheur}, {Bernath}, {Jorissen}, \&
  {Christlieb}}]{masseron}
{Masseron}, T., {Plez}, B., {Van Eck}, S., {et~al.} 2014, \aap, 571, A47

\bibitem[{{McWilliam} \& {Lambert}(1988)}]{mcwilliam}
{McWilliam}, A. \& {Lambert}, D.~L. 1988, \mnras, 230, 573

\bibitem[{{Mel{\'e}ndez} \& {Cohen}(2009)}]{melendezm71}
{Mel{\'e}ndez}, J. \& {Cohen}, J.~G. 2009, \apj, 699, 2017

\bibitem[{{Milone} {et~al.}(2012){Milone}, {Piotto}, {Bedin}, {King},
  {Anderson}, {Marino}, {Bellini}, {Gratton}, {Renzini}, {Stetson}, {Cassisi},
  {Aparicio}, {Bragaglia}, {Carretta}, {D'Antona}, {Di Criscienzo},
  {Lucatello}, {Monelli}, \& {Pietrinferni}}]{milone}
{Milone}, A.~P., {Piotto}, G., {Bedin}, L.~R., {et~al.} 2012, \apj, 744, 58

\bibitem[{{Nomoto} {et~al.}(2006){Nomoto}, {Tominaga}, {Umeda}, {Kobayashi}, \&
  {Maeda}}]{nomoto}
{Nomoto}, K., {Tominaga}, N., {Umeda}, H., {Kobayashi}, C., \& {Maeda}, K.
  2006, Nuclear Physics A, 777, 424

\bibitem[{{Origlia} {et~al.}(2011){Origlia}, {Rich}, {Ferraro}, {Lanzoni},
  {Bellazzini}, {Dalessandro}, {Mucciarelli}, {Valenti}, \&
  {Beccari}}]{2011ApJ...726L..20O}
{Origlia}, L., {Rich}, R.~M., {Ferraro}, F.~R., {et~al.} 2011, \apjl, 726, L20

\bibitem[{{Plez}(1998)}]{plez}
{Plez}, B. 1998, \aap, 337, 495

\bibitem[{{Prantzos} {et~al.}(2007){Prantzos}, {Charbonnel}, \&
  {Iliadis}}]{prantzos2}
{Prantzos}, N., {Charbonnel}, C., \& {Iliadis}, C. 2007, \aap, 470, 179

\bibitem[{{Prantzos} \& {Goswami}(2001)}]{prantzos}
{Prantzos}, N. \& {Goswami}, A. 2001, Nuclear Physics A, 688, 37

\bibitem[{{Ram{\'{\i}}rez} \& {Allende Prieto}(2011)}]{ramirez}
{Ram{\'{\i}}rez}, I. \& {Allende Prieto}, C. 2011, \apj, 743, 135

\bibitem[{{Ram{\'{\i}}rez} {et~al.}(2008){Ram{\'{\i}}rez}, {Allende Prieto},
  {Lambert}, \& {Asplund}}]{ramirez2}
{Ram{\'{\i}}rez}, I., {Allende Prieto}, C., {Lambert}, D.~L., \& {Asplund}, M.
  2008, in ASPCS, Vol. 393, New Horizons in Astronomy, ed. A.~{Frebel}, J.~R.
  {Maund}, J.~{Shen}, \& M.~H. {Siegel}, 255

\bibitem[{{Sbordone} {et~al.}(2014){Sbordone}, {Caffau}, {Bonifacio}, \&
  {Duffau}}]{sbordone14}
{Sbordone}, L., {Caffau}, E., {Bonifacio}, P., \& {Duffau}, S. 2014, \aap, 564,
  A109

\bibitem[{{Shayesteh} \& {Bernath}(2011)}]{shayesteh}
{Shayesteh}, A. \& {Bernath}, P.~F. 2011, \jcp, 135, 094308

\bibitem[{{Shetrone}(1996)}]{shetrone}
{Shetrone}, M.~D. 1996, \aj, 112, 2639

\bibitem[{{Sneden} {et~al.}(2012){Sneden}, {Bean}, {Ivans}, {Lucatello}, \&
  {Sobeck}}]{sneden2}
{Sneden}, C., {Bean}, J., {Ivans}, I., {Lucatello}, S., \& {Sobeck}, J. 2012,
  {MOOG: LTE line analysis and spectrum synthesis}, astrophysics Source Code
  Library

\bibitem[{{Sneden} {et~al.}(2014){Sneden}, {Lucatello}, {Ram}, {Brooke}, \&
  {Bernath}}]{snedencn}
{Sneden}, C., {Lucatello}, S., {Ram}, R.~S., {Brooke}, J.~S.~A., \& {Bernath},
  P. 2014, \apjs, 214, 26

\bibitem[{{Sneden}(1973)}]{sneden1}
{Sneden}, C.~A. 1973, PhD thesis, The University of Texaz at Austin.

\bibitem[{{Sobeck} {et~al.}(2011){Sobeck}, {Kraft}, {Sneden}, {Preston},
  {Cowan}, {Smith}, {Thompson}, {Shectman}, \& {Burley}}]{sobeck}
{Sobeck}, J.~S., {Kraft}, R.~P., {Sneden}, C., {et~al.} 2011, \aj, 141, 175

\bibitem[{{Thygesen} {et~al.}(2014){Thygesen}, {Sbordone}, {Andrievsky},
  {Korotin}, {Yong}, {Zaggia}, {Ludwig}, {Collet}, {Asplund}, {Ventura},
  {D'Antona}, {Mel{\'e}ndez}, \& {D'Ercole}}]{thygesen}
{Thygesen}, A.~O., {Sbordone}, L., {Andrievsky}, S., {et~al.} 2014, \aap, 572,
  A108

\bibitem[{{Tominaga} {et~al.}(2007){Tominaga}, {Umeda}, \& {Nomoto}}]{tominaga}
{Tominaga}, N., {Umeda}, H., \& {Nomoto}, K. 2007, \apj, 660, 516

\bibitem[{{Ventura} {et~al.}(2011){Ventura}, {Carini}, \&
  {D'Antona}}]{venturaMgAl2011}
{Ventura}, P., {Carini}, R., \& {D'Antona}, F. 2011, \mnras, 415, 3865

\bibitem[{{Ventura} {et~al.}(2014){Ventura}, {Criscienzo}, {D'Antona},
  {Vesperini}, {Tailo}, {Dell'Agli}, \& {D'Ercole}}]{ventura}
{Ventura}, P., {Criscienzo}, M.~D., {D'Antona}, F., {et~al.} 2014, \mnras, 437,
  3274

\bibitem[{{Ventura} \& {D'Antona}(2008)}]{ventura2008}
{Ventura}, P. \& {D'Antona}, F. 2008, \mnras, 385, 2034

\bibitem[{{Ventura} \& {D'Antona}(2009)}]{ventura2009}
{Ventura}, P. \& {D'Antona}, F. 2009, \aap, 499, 835

\bibitem[{{Ventura} \& {D'Antona}(2011)}]{ventura2011hbb}
{Ventura}, P. \& {D'Antona}, F. 2011, \mnras, 410, 2760

\bibitem[{{Villanova} {et~al.}(2013){Villanova}, {Geisler}, {Carraro}, {Moni
  Bidin}, \& {Mu{\~n}oz}}]{villanova}
{Villanova}, S., {Geisler}, D., {Carraro}, G., {Moni Bidin}, C., \&
  {Mu{\~n}oz}, C. 2013, \apj, 778, 186

\bibitem[{{Wallerstein} {et~al.}(1997){Wallerstein}, {Iben}, {Parker},
  {Boesgaard}, {Hale}, {Champagne}, {Barnes}, {K{\"a}ppeler}, {Smith},
  {Hoffman}, {Timmes}, {Sneden}, {Boyd}, {Meyer}, \& {Lambert}}]{wallerstein}
{Wallerstein}, G., {Iben}, Jr., I., {Parker}, P., {et~al.} 1997, Reviews of
  Modern Physics, 69, 995

\bibitem[{{Williams} {et~al.}(2010){Williams}, {Bureau}, \&
  {Cappellari}}]{williams}
{Williams}, M.~J., {Bureau}, M., \& {Cappellari}, M. 2010, \mnras, 409, 1330

\bibitem[{{Yong} {et~al.}(2006){Yong}, {Aoki}, \& {Lambert}}]{yongm71}
{Yong}, D., {Aoki}, W., \& {Lambert}, D.~L. 2006, \apj, 638, 1018

\bibitem[{{Yong} {et~al.}(2003{\natexlab{a}}){Yong}, {Grundahl}, {Lambert},
  {Nissen}, \& {Shetrone}}]{yong6752}
{Yong}, D., {Grundahl}, F., {Lambert}, D.~L., {Nissen}, P.~E., \& {Shetrone},
  M.~D. 2003{\natexlab{a}}, \aap, 402, 985

\bibitem[{{Yong} {et~al.}(2003{\natexlab{b}}){Yong}, {Lambert}, \&
  {Ivans}}]{yong_field}
{Yong}, D., {Lambert}, D.~L., \& {Ivans}, I.~I. 2003{\natexlab{b}}, \apj, 599,
  1357

\end{thebibliography}
